\documentclass{aa}

\usepackage{natbib}
\bibpunct{(}{)}{;}{a}{}{,}
\usepackage[utf8]{inputenc}
\usepackage[varg]{txfonts}
\usepackage{amsmath}
\usepackage{graphicx}
\usepackage{lscape}
\usepackage{multirow}
\usepackage{xcolor}
\usepackage{dcolumn}
\newcolumntype{d}[1]{D{.}{.}{#1}}

\newcommand{\sci}[2]{\ensuremath{#1 \times 10^{#2}}}
\newcommand{\kms}[1]{#1\,km\,s\ensuremath{^{-1}}}
\newcommand{\mjyb}[1]{#1\,mJy\,beam\ensuremath{^{-1}}}

\newcommand{\uncsci}[4]{\ensuremath{\left(#1_{#2}^{#3}\right) \times 10^{#4}}}
\newcommand{\unc}[3]{\ensuremath{#1_{#2}^{#3}}}

\usepackage[colorlinks=true, linkcolor=blue, citecolor=blue, filecolor=blue, urlcolor=blue]{hyperref}

\begin{document}
\title{The hot corino-like chemistry of four FUor-like protostars}

\author{F.~Cruz-Sáenz de Miera\inst{1}\fnmsep\inst{2}
        \and
        A.~Coutens\inst{1}
        \and
        Á.~Kóspál\inst{2}\fnmsep\inst{3}\fnmsep\inst{4}
        \and
        P.~Ábrahám\inst{2}\fnmsep\inst{3}
        \and
        A.~Dutrey\inst{5}
        \and
        S.~Guilloteau\inst{5}
       }

\institute{Institut de Recherche en Astrophysique et Plan\'etologie, Universit\'e de Toulouse, UT3-PS, OMP, CNRS, 9 av.\ du Colonel Roche, 31028 Toulouse Cedex 4, France
           \email{fernando.cruz-saenz $<$at$>$ irap.omp.eu}
           \and
           Konkoly Observatory, HUN-REN Research Centre for Astronomy and Earth Sciences, MTA Centre of Excellence, Konkoly-Thege Mikl\'os \'ut 15-17, 1121 Budapest, Hungary
           \and
           ELTE E\"otv\"os Lor\'and University, Institute of Physics, P\'azm\'any P\'eter s\'et\'any 1/A, 1117 Budapest, Hungary
           \and
           Max Planck Institute for Astronomy, Königstuhl 17, 69117 Heidelberg, Germany
           \and
           Laboratoire d’astrophysique de Bordeaux, Univ. Bordeaux, CNRS, B18N, allée Geoffroy Saint-Hilaire, 33615 Pessac, France
          }

\abstract
{
Compared to Class~0 protostars, the higher densities and lower temperatures of the disk midplanes of Class~I young stellar objects (YSOs) limit the detectability of complex organic molecules (COMs). The elevated luminosities of eruptive YSOs increase disk temperatures sublimating frozen molecules and easing their detection.
}
{
Our aim is to investigate the chemical composition of four FUor-like Class I YSOs: L1551~IRS~5, Haro~5a~IRS, V346~Nor, and OO~Ser, and to compare their abundances of COMs with other YSOs in the literature.
}
{
We search for COMs line emission in ALMA Band 6 observations. We use the CASSIS software to determine their column densities ($N$) and excitation temperatures ($T_\mathrm{ex}$) assuming local thermodynamical equilibrium.
}
{
We detect 249 transitions from 12 COMs. In L1551~IRS~5 we identified CH\(_3\)OH, \(^{13}\)CH\(_3\)OH, CH\(_3\)\(^{18}\)OH, CH\(_2\)DOH, CH\(_3\)CHO, CH\(_3\)OCH\(_3\), CH\(_3\)OCHO, CH\(_3\)COCH\(_3\), C\(_2\)H\(_5\)OH, C\(_2\)H\(_5\)CN, \(^{13}\)CH\(_3\)CN, and CH\(_3\)C\(^{15}\)N. Haro~5a~IRS and OO~Ser have emission from CH\(_3\)OH, CH\(_3\)CHO, CH\(_3\)OCH\(_3\), and CH\(_3\)OCHO. CH\(_3\)COCH\(_3\) is also detected in OO~Ser. In V346~Nor we found CH\(_3\)OH, CH\(_2\)DOH, CH\(_3\)CHO, CH\(_3\)OCH\(_3\), CH\(_3\)OCHO, and C\(_2\)H\(_5\)CN.
The emission of COMs is compact in all targets.
The analysis indicates their temperatures are above 100\,K.
The abundance ratios of COMs derived for these eruptive YSOs, as well as for other protostars in the literature, span several orders of magnitude without any clear differentiation between the eruptive and quiescent YSOs.
The column density of the main isotopologue of CH\(_3\)OH should not be used as a reference, as most of the lines are optically thick.
}
{
The hot and compact emission of COMs indicates that the four FUor-like targets are hot corino-like. Spectral studies of such objects can be useful to investigate the complex organic chemistry at later evolutionary stages than the usual Class~0 stage.
}


\maketitle

\nolinenumbers
\section{Introduction}\label{sec:intro}
Recent results have proposed that planets start forming in the early stages of protostellar evolution, when the young stellar objects (YSOs) are still deeply embedded in their envelopes \citep[e.g.,][]{Drazkowska2023_ASPC534717D}.
The chemical composition of these planets is expected to be inherited from the protoplanetary disks in which they are forming.
Part of this composition is dictated by the presence of complex organic molecules (COMs).
These are molecules composed of at least 6 atoms, of which at least one is carbon, and are the precursors of more complex pre-biotic species \citep{Herbst2009_ARAA47427H}.
During the early stages of the star formation process, when the temperatures are more elevated due to the higher mass accretion rates, COMs are more easily detected as they are still in the gas phase.
In the case of low-mass protostars, the compact regions in which COMs are abundant in the gas phase are called hot corinos \citep{Ceccarelli2004_ASPC323195C}.
Hot corinos have been detected in both Class~0 and Class~I sources \citep[e.g.,][]{Cazaux2003_ApJ593L51C,Bottinelli2004_ApJ617L69B,Bergner2019_ESC31564B,Bianchi2019_MNRAS4831850B}.
However, the more evolved Class~I protostars have lower COM detection rates \citep{ArturdelaVillarmois2019_AA626A71A,Yang2022_ApJ92593Y}.
This lack of detections can be attributed to a combination of a lower temperature, a smaller hot region and a lower envelope density \citep{vantHoff2020_ApJ901166V,Hsu2023_ApJ956120H}.

FU~Orionis-type stars (FUors) are young stellar objects that experience a sudden increase in their mass accretion rate 2 to 5 orders of magnitude, reaching values as high as 10$^{-4}$\,M$_\odot$\,year$^{-1}$.
\citet{Fischer2022_PP7} presented the most recent review of eruptive young stars, including FUors.
This surge of the mass accretion rate is reflected as an increase in the accretion luminosity such that the bolometric luminosity of the protostar is then dominated by it.
During this state of elevated accretion luminosity, the temperatures within $\sim$10\,au from the protostar rise to $>$100\,K \citep[e.g., FU~Ori and V883~Ori;][]{Lykou2022_AA663A86L,Bourdarot2023_AA676A124B,Alarcon2024_MNRAS5279655A}, so that the protostellar disks and the envelope undergo detectable changes in their physical, mineralogical and chemical properties \citep[e.g.,][]{Molyarova2018_ApJ86646M,Tobin2023_Natur615227T,Zwicky2024_MNRAS5277652Z}.
In particular, the increase in temperature causes the desorption of molecules that were frozen on the dust grains in the system.

Theoretical works have studied the effects of these outbursts, including the estimation of the changes in the abundances and distribution of molecules throughout the FUor envelopes and disks.
\citet{Taquet2016_ApJ82146T} studied the changes in the abundances of dimethyl ether (CH$_3$OCH$_3$) and methyl formate (CH$_3$OCHO) out to envelope scales, and found that during a FUor-type outburst the abundances of these molecules with respect to (w.r.t.) methanol (CH\(_3\)OH) should be a factor of a few higher than pre- or post-outburst.
\citet{Molyarova2018_ApJ86646M} modeled the changes in the protostellar disk during and after a FUor-type outburst, and found that the abundances of several molecules (including some COMs) can increase by $\sim$3 to 8 orders of magnitude during the outburst and remain elevated between 20 and 120 years after the protostar has returned to its quiescent state.
\citet{Zwicky2024_MNRAS5277652Z} carried out a similar study to estimate the impact of these powerful accretion events and found a sub-set of molecules (e.g., N$_2$H$^+$, C$^{18}$O, H$_2$CO and HCN) that can be used as tracers of previous outbursts.

Interferometric observations have resulted in detections of multiple COMs in the disks of five eruptive young stars:
\object{SVS13-A} \citep{Bianchi2017_MNRAS4673011B,Bianchi2019_MNRAS4831850B,Bianchi2022_ApJ928L3B},
\object{V883~Ori} \citep{vantHoff2018ApJ864L23V,Lee2019NatAs3314L,Yamato2024_AJ16766Y},
L1551~IRS~5 \citep{Bianchi2020_MNRAS498L87B},
\object{HOPS~373W} \citep{Lee2023_ApJ95643L},
and \object{B335} \citep{Lee2025_ApJ978L3L}.
\object{SVS13-A} (also known as V512~Per) is a Class~I YSO that underwent a $\Delta K\sim$1.5\,mag brightening in 1990 and has been fading since then \citep{Eisloeffel1991_ApJ383L19E,Hodapp2014_ApJ794169H}.
Based on the amplitude of its brightening and on its optical/near-infrared spectrum, it was first classified as an EX~Lupi-type outbursting young star (EXor).
However, the length of the accretion outburst ($>$30\,years) is vastly longer than what is expected in EXors ($\lesssim$1\,year) and instead its duration is more comparable to the one of FUor-type outbursts.
The COMs detected around \object{SVS13-A} are CH\(_3\)CHO, CH\(_3\)OCHO, CH\(_3\)OCH\(_3\), C\(_2\)H\(_5\)OH, NH$_2$CHO, CH\(_3\)CN, CH\(_2\)DCN, and HCOCH$_2$OH \citep{DeSimone2017_AA599A121D,Bianchi2019_MNRAS4831850B,Bianchi2022_AA662A103B,Bianchi2022_ApJ928L3B}.
\object{V883~Ori} is a Class~I young eruptive star located in Orion classified as FUor-like.
This classification means that the brightening due to the accretion outburst was not detected, but the protostar presents the near-infrared spectroscopic signatures typical of FUors \citep{ConnelleyReipurth2018_ApJ861145C}.
The molecules detected in \object{V883~Ori} are CH\(_3\)OH, CH$_2$DOH, \(^{13}\)CH\(_3\)OH, CH$_3$CHO, $^{13}$CH\(_3\)CHO, CH$_3$OCHO, CH\(_3\)O$^{13}$CHO, $^{13}$CH\(_3\)OCHO, CH\(_2\)DOCHO, CH\(_3\)OCDO, CH\(_3\)OCH\(_3\), CH$_3$COCH$_3$, CH$_3$CN, $c$-C$_2$H$_4$O, $t$-HCOOH, $t$-C$_2$H$_3$CHO, $s$-C$_2$H$_5$CHO, and CH$_3$SH\@ \citep{vantHoff2018ApJ864L23V,Lee2019NatAs3314L,Yamato2024_AJ16766Y}.
HOPS~373W is a young protostar in Orion that experienced a brief outburst in 2019 and began another accretion event in 2020 \citep{Yoon2022_ApJ92960Y}.
\citet{Lee2023_ApJ95643L} observed HOPS~373W with Band 7 of the Atacama Large Millimeter Array (ALMA).
They identified multiple COMs including CH\(_3\)OH and several deuterated forms of it, CH\(_3\)CHO, CH\(_3\)OCHO, C\(_2\)H\(_5\)OH, CH\(_3\)CN, and HCOCH$_2$OH.
B335 is a Class~0 protostar that experienced a burst between 2010/2012 and 2023, detected only at mid- and far-infrared wavelengths.
\citet{Lee2025_ApJ978L3L} examined four epochs of ALMA data to study the changes in the column densities of \(^{13}\)CH\(_3\)OH, CH\(_2\)DOH, CH\(_3\)OCHO, C\(_2\)H\(_5\)OH, HC(O)NH$_2$, and DC(O)NH$_2$.
L1551~IRS~5 is one of the FUor-type eruptive young stars that we focus on during this paper and is described in detail below.

In this paper, we present a chemical study of four FUor-like eruptive young stars: L1551~IRS~5, Haro~5a~IRS, OO~Ser, and V346~Nor, using ALMA Band~6 observations.
L1551~IRS~5 is a Class~I protobinary system located in the Taurus star-forming region at a distance of 147\,pc from the Sun with a bolometric luminosity ($L_\mathrm{bol}$) of 40\,L$_\odot$ \citep{Liseau2005_ApJ619959L}.
The Northern star is more massive and luminous that the Southern one \citep{Liseau2005_ApJ619959L}.
The separation between the two protostars is 50\,au.
Their circumstellar disks have radii $<$10\,au with both disks being more massive than those of typical young stellar objects \citep{Lim2016ApJ826153L,CruzSaenzdeMiera2019_ApJ882L4C,Kospal2021_ApJS25630K}.
L1551~IRS~5 also contains a $\sim$140\,au circumbinary ring which is less massive than the two protostellar disks and has a slightly different inclination angle than them \citep{Lim2016ApJ826153L,CruzSaenzdeMiera2019_ApJ882L4C,Takakuwa2020_ApJ89810T}.
Using ALMA Band~6 observations, \citet{Bianchi2020_MNRAS498L87B} detected three COMs toward this target: methanol (CH$_3$OH, $^{13}$CH$_3$OH, and CH$_2$DOH), methyl formate (HCOOCH$_3$), and ethanol (CH$_3$CH$_2$OH).
Their beam size was not sufficient to resolve the two components of the system.
However, using the methanol emission, they found differences in the systemic velocities ($\sim$\kms{3}) along the length of the system in a North-to-South direction.
\citet{Andreu2023_AA677L17A} observed HDO and H\(_2\)\(^{18}\)O in L1551~IRS~5, and while their Northern Extended Millimetre Array (NOEMA) observations did not resolve the protobinary either, they showed a comparable difference in velocities.
\citet{Takakuwa2020_ApJ89810T} resolved the protobinary using ALMA Band 7 observations of C$^{18}$O.
They confirmed that the difference in velocities is due to a combination of the difference in radial velocities between the two protostars and the velocity gradient due to the disks rotations.
\citet{Mercimek2022_AA659A67M} observed L1551~IRS~5 using the Institut de radioastronomie millimétrique (IRAM) 30\,m antenna, and covering two spectral windows: one between 214.5 and 222.2\,GHz, and another between 230.2 and 238.0\,GHz (i.e., $\sim$1\,mm in wavelength).
They identified 33 different species in the spectrum of L1551~IRS~5, including six COMs: CH\(_3\)OH, CH\(_2\)DOH, CH\(_3\)CN, CH\(_3\)CCH, CH\(_3\)CHO and CH\(_3\)OCHO.
\citet{Marchand2024_AA687A195M} also used the 30\,m telescope to observe L1551~IRS~5 at 2\,mm and 3\,mm.
The authors identified emission lines of 75 different species, including 10 COMs: CH$_3$OH, CH$_2$DOH, CH$_3$CHO, CH$_3$OCH$_3$, CH$_3$OCHO, CH$_3$CN, $l$-C$_4$H$_2$, CH$_3$CCH, CH$_2$DCCH, and HC$_5$N.
Although their angular resolution was not enough to resolve the protostellar disks from the envelope, they found that some species have two components at two different temperatures (10\,K and $>$30\,K) which would trace the cold envelope and the warmer emission close to the protostars.

For the other three FUor-like sources, this is the first study delving into the chemistry of their disks.
Haro~5a~IRS is a Class~I protostar at a distance of 391\,pc in Orion with $L_\mathrm{bol}$ = 50\,L$_\odot$ \citep{Reipurth1997_AJ1142700R}, and it has a weaker companion with a separation of 0.70$''$ to the Southwest \citep{Tobin2020_ApJ890130T,Kospal2021_ApJS25630K}.
Haro~5a~IRS and its companion have dust disk radii of 39\,au and 34\,au \citep{Kospal2021_ApJS25630K}.
V346~Nor is a Class~0/I in the Norma star-forming region with a distance of 700\,pc with $L_\mathrm{bol}$ = 160\,L$_\odot$ \citep{Kospal2017_ApJ84345K}, whose outburst began before 1978.
In 2010 V346~Nor dimmed suddenly, however, after a few months in quiescence, it began to brighten again, indicating the onset of a new outburst \citep{Kraus2016_MNRAS462L61K,Kospal2020_ApJ889148K}.
OO~Ser is a Class~I protostar located in the Serpens NW star-forming region, with a distance of 438\,pc.
Its outburst started between 1991 and 1994, and it began dimming after reaching its peak brightness in late 1995 \citep{Kospal2007_AA470211K,Hodapp2012_ApJ74456H}.
\citet{Kospal2007_AA470211K} calculated the pre-outburst and outburst bolometric luminosities of OO~Ser, which, after we adjusted them to the assumed distance of 438\,pc, are 9 and 70\,L$_\odot$, respectively.
ALMA observations revealed that its dust disk has radius of 26\,au \citep{Kospal2021_ApJS25630K}.

We present the ALMA observations and the data reduction in Sect.~\ref{sec:observations}.
The resulting spectra and moment maps are shown in Sect.~\ref{sec:results}.
In Sect.~\ref{sec:analysis} and Sect.~\ref{sec:discussion} we present the analysis and the discussion.
Finally, our conclusions are in Sect.~\ref{sec:conclusions}.

\section{Observations}\label{sec:observations}
The four FUor-like sources were observed as part of the ALMA program 2016.1.00209.S\@ (PI\@: Takami) in Cycle 4.
The four targets were observed using ALMA's 7m array plus two extended configurations (C40-3 and C40-6) resulting in baseline ranges between 8\,m and $\sim$3650\,m.
The four targets had similar integration times.
For L1551~IRS~5 these were 4.6, 3.6, and 11.3 minutes for the 7m, C40-3 and C40-6 arrays, respectively.
In the case of Haro~5a~IRS these were 4.1, 3.1 and 9.3 minutes, respectively, with the 7m and C40-6 observations being repeated due to weather conditions.
For V346~Nor and OO~Ser the integration times were 4.1, 3.1 and 9.8 minutes for the three arrays.
In the case of OO~Ser, the 7m array observations were also repeated.
The dates of the observations and other details can be found in Table~1 of \citet{Kospal2021_ApJS25630K}.
Data were taken in Band 6, with three spectral windows centered at the $^{12}$CO, $^{13}$CO, and C$^{18}$O $J=2-1$ lines, and two spectral windows of 1875 MHz bandwidth each, centered at 216.9\,GHz and 232.2\,GHz.
These latter two spectral windows are the focus of this work.
They were taken at a spectral resolution of $\sim$0.490\,MHz, i.e., \kms{0.677} and \kms{0.633}.

The data sets were calibrated using Common Astronomy Software Applications (CASA) v.5.1.1 \citep{McMullin2007ASPC376127M}.
We carried out self-calibration for both the visibility phases and amplitudes to improve the signal-to-noise ratio (SNR), and we applied these solutions to the continuum-subtracted measurement sets.
The common baselines between the three different observing configurations were consistent so we concatenated them into a single measurement set.
This concatenated measurement set includes observations using the full baseline range and thus recovers the emission at most spatial scales.

The continuum-subtracted data cubes were created using \textit{tclean} in CASA with a Briggs scheme and a robust value of 0.5, thus generating maps with a compromise between sensitivity and angular resolution.
Both data cubes were made using the same image size, pixel size, and beam shape that was obtained using the spectral window with the shortest frequencies.
The beam shapes for each target and the resulting rms for each spectral window can be found in \autoref{tab:observations}.

\begin{table}
  \caption{Beam shapes and position angles for each target, and the rms calculated for each spectral window and a spectral resolution of 0.490\,MHz.\label{tab:observations}}
  \begin{center}
    \begin{tabular}[c]{ccccc}
      \hline\\[-9pt]
      Target      & Beam size          & Beam PA    & rms$_{216}$ & rms$_{232}$\\
                  & [$''$]             & [$^\circ$] & [K]         & [K]\\
      \hline\\[-9pt]
      L1551~IRS~5 & 0.204$\times$0.189 & $-$60.8    & 1.35        & 1.27\\
      Haro~5a~IRS & 0.278$\times$0.141 & $-$78.2    & 1.35        & 1.27\\
      V346~Nor    & 0.171$\times$0.116 & $+$87.3    & 1.34        & 1.30\\
      OO~Ser      & 0.221$\times$0.152 & $-$68.3    & 1.35        & 1.27\\
      \hline
    \end{tabular}
  \end{center}
\end{table}

\section{Results}\label{sec:results}
In this Section we present the extraction of the spectra used in our analysis, the identification of the emission lines, and discuss the morphology of the emission of the COMs and the smaller molecules.

\subsection{Spectra}\label{ss:spectra}
\begin{figure*}
\centering
\includegraphics[width=0.2475\linewidth]{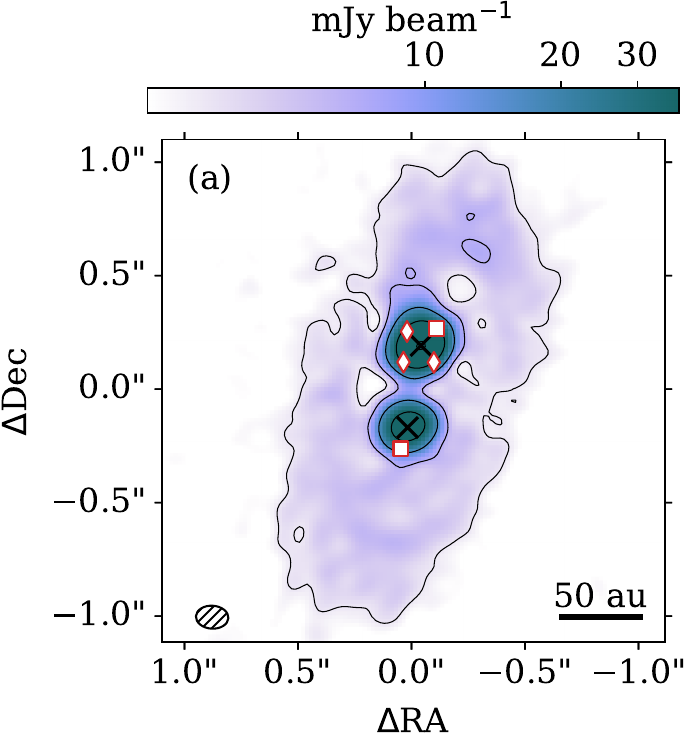}\hfill
\includegraphics[width=0.24\linewidth]{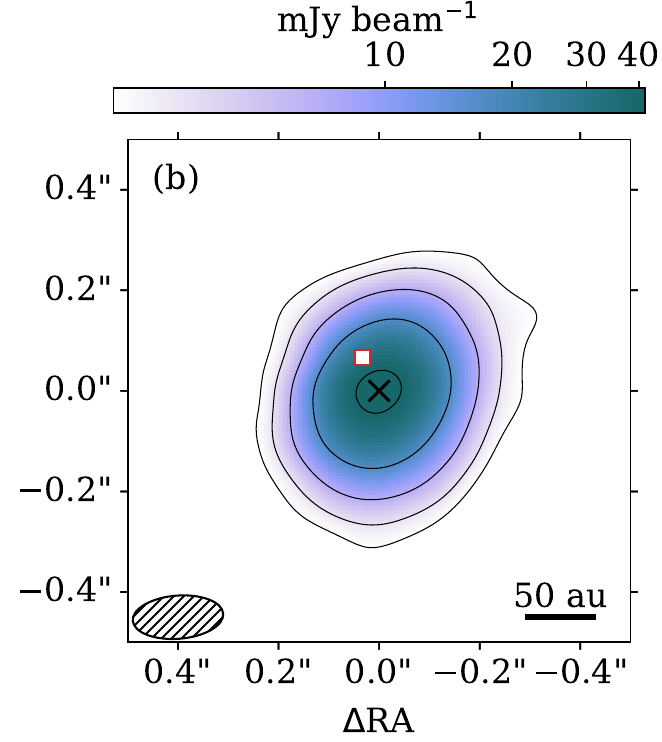}\hfill
\includegraphics[width=0.24\linewidth]{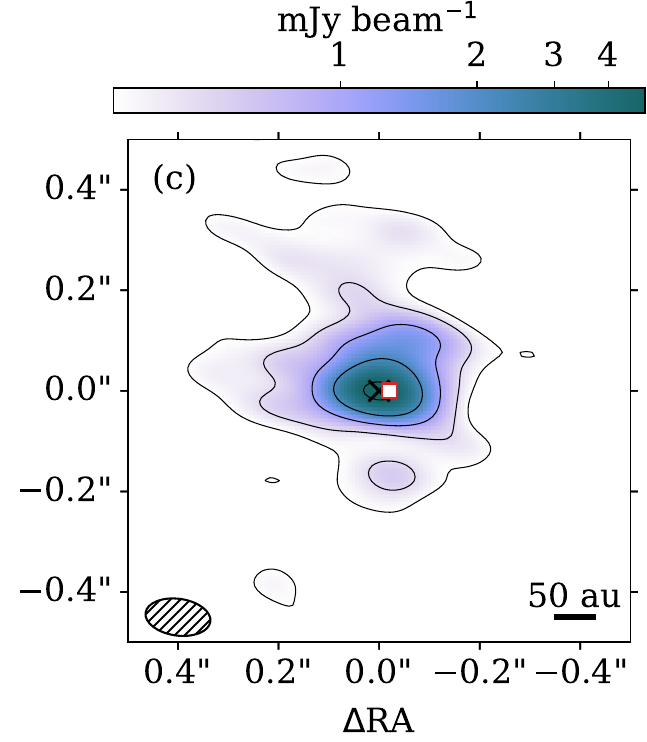}\hfill
\includegraphics[width=0.24\linewidth]{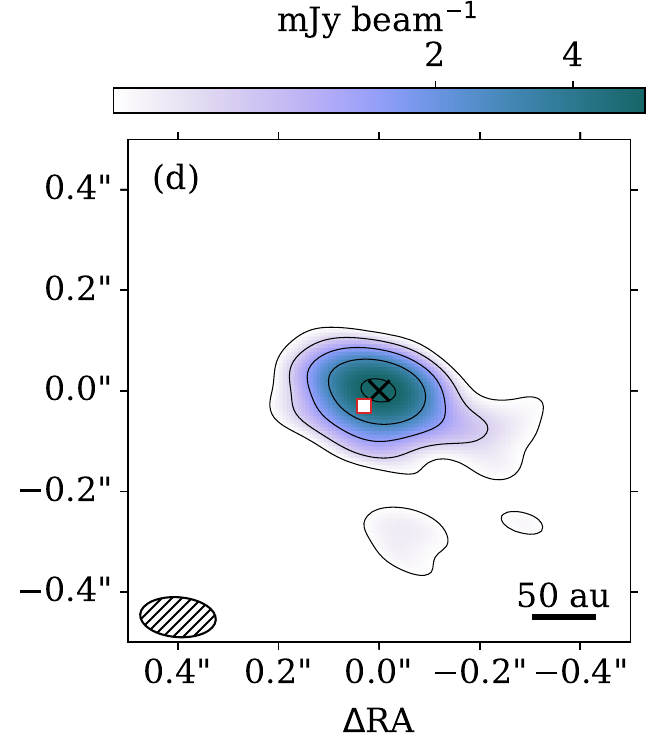}
\caption{Continuum maps of the four FUor sources at 1.33\,mm.
In all panels, the cross markers indicate the continuum peaks, i.e., the positions of the protostars, and the square markers the pixels used to extract the spectra.
In the case of L1551~IRS~5, the diamond markers indicate the positions of the alternative pixels use to examine potential spatial changes of the abundances (see Section\,\ref{ss:impact_other_pixels}).
Panel \textit{(a)} shows the map of L1551~IRS~5 centered on the position between the two protostars. The contours are at 8, 18, 44, 105 and 250\(\sigma\) with \(\sigma\) = \mjyb{0.37}.
Panel \textit{(b)} is the map of Haro~5a~IRS with contours at 5, 12, 30, 75 and 186\(\sigma\) with \(\sigma\) = \mjyb{0.22}.
Panel \textit{(c)} is V346~Nor with contours at 5, 10, 23, 49 and 106\(\sigma\) with \(\sigma\) = \mjyb{0.05}.
Panel \textit{(d)} shows OO~Ser with contours at 5, 12, 30, 75 and 186\(\sigma\) with \(\sigma\) = \mjyb{0.22}.}
\label{fig:pixels}
\end{figure*}

In their analysis of the continuum, \citet{Kospal2021_ApJS25630K} found that the inner disks of our targets are optically thick.
Thus, line emission is expected to be weaker close to the peak continuum position.
We verified this by extracting spectra from the pixel that matches the position of the peak in the continuum maps, i.e.\ the position of the protostar, and at some random pixels separated from the center.
We found weaker line emission in the spectra extracted on the pixel closest to the protostar.
Thus, we decided to extract the spectra from off-center positions where we could find the strongest line emission.
We chose these positions by extracting spectra using a pixel-size aperture placed at multiple positions throughout the disks, and finding the pixels where the line emission was stronger in both spectral windows.
Our decision to use an individual pixel to extract the spectra, instead of a larger aperture to boost the SNR, was to minimize the blending caused by the distribution of velocities found in a larger aperture.
The continuum maps showing the positions of the protostars and our apertures are shown in \autoref{fig:pixels}.
The coordinates of all the protostars and the ones of the selected positions for the extraction of spectra can be found in \autoref{tab:positions}.
For all targets, the separation between the peak of the continuum and where we extracted the spectra is less than the beam size.

\begin{table*}
  \caption{Positions of the protostars and of the pixels used to extract the spectra, and the separation between both positions.}
  \label{tab:positions}
  \begin{center}
    \begin{tabular}[c]{ccccd{1.2}}
      \hline\\[-9pt]
      Target        & Protostar coordinates         & Spectra coordinates           & Separation & \multicolumn{1}{c}{$v_\mathrm{sys}$}\\
                    &                               &                               & [$''$]     & \multicolumn{1}{c}{[km\,s$^{-1}$]}\\
      \hline\\[-9pt]
      L1551~IRS~5 N & 04:31:34.1623 $+$18:08:04.717 & 04:31:34.1575 $+$18:08:04.793 & 0.102      & 7.56\\
      L1551~IRS~5 S & 04:31:34.1664 $+$18:08:04.356 & 04:31:34.1669 $+$18:08:04.253 & 0.103      & 5.16\\
      Haro~5a~IRS   & 05:35:26.5594 $-$05:03:55.130 & 05:35:26.5616 $-$05:03:55.064 & 0.074      & 14.06\\
      V346~Nor      & 16:32:32.1971 $-$44:55:30.769 & 16:32:32.1951 $-$44:55:30.769 & 0.021      & -4.58\\
      OO~Ser        & 18:29:49.1515 $+$01:16:19.676 & 18:29:49.1535 $+$01:16:19.646 & 0.042      & 8.22\\
      \hline
    \end{tabular}
    \tablefoot{The $v_\mathrm{LSR}$ at the coordinates where we extracted the spectra is measured on the bright H$_2$S 2(2,0)-2(1,1) transition at 216.71\,GHz detected in these observations. }
  \end{center}
\end{table*}

For the line identification we used the CASSIS\footnote{CASSIS has been developed by IRAP-UPS/CNRS \citep{Vastel2015_sf2a.conf..313V}. \url{http://www.cassis.irap.omp.eu/}} software, which includes the Cologne Database for Molecular Spectroscopy\footnote{\url{https://cdms.astro.uni-koeln.de/}} \citep[CDMS;]
[]{Muller2001_AA370L49M,Muller2005_JMoSt742215M}, the Jet Propulsion Laboratory \citep[JPL;][]{Pickett1998_JQSRT60883P} molecular database\footnote{\url{https://spec.jpl.nasa.gov/}} and the Lille Spectroscopic Database\footnote{\url{https://lsd.univ-lille.fr/}} (LSD).
We used CDMS for most of the species.
For CHD\(_2\)OH, CH\(_3\)COCH\(_3\), and CH\(_3\)OCHO, we utilized the JPL database, and for CH\(_3\)CHO we used LSD.
The spectroscopic references for the different molecules are: 
CH\(_3\)OH \citep{Pickett1981_JMoSp89542P,Sastry1984_JMoSp103486S,Herbst1984_JMoSp10842H,Anderson1990_ApJS72797A,Odashima1995_JMoSp173404O,Muller2004_AA428.1019M,Xu2008_JMoSp251305X},
\(^{13}\)CH\(_3\)OH \citep{Xu1996_JMoSp179269X,Xu1997_JPCRD2617X},
CH\(_3\)\(^{18}\)OH \citep{Fisher2007_JMoSp2457F},
CH$_2$DOH \citep{Pearson2012_JMoSp280119P},
CH$_3$CHO \citep{Smirnov2014_JMoSp29544S},
CH\(_3\)OCH\(_3\) \citep[and references therein]{Endres2009_AA504635E},
CH$_3$OCHO \citep[and references therein]{Ilyushin2009_JMoSp25532I},
CH$_3$COCH$_3$ \citep[and references therein]{Oldag1992_ZNatA47527O,Groner2002_ApJS142145G},
C$_2$H$_5$OH \citep{Pearson1995_JPCRD241P,Pearson1996_JMoSp175246P,Pearson2008_JMoSp251394P,Muller2016_AA587A92M},
C$_2$H$_5$CN \citep[and references therein]{Johnson1977_ApJ218370J,Boucher1980_ZNatA351136B,Pearson1994_ApJS93589P,Fukuyama1996_ApJS104329F,Brauer2009_ApJS184133B}, and
$^{13}$CH$_3$CN and CH$_3$C{$^{15}$N} \citep[and references therein]{Muller2009_AA5061487M}.

Among our targets, L1551~IRS~5 is the one with the most detected species, which we suspect is due to a combination of its proximity and its luminosity.
In its Northern disk we identified simple organics (H\(_2\)CO, D\(_2\)CO, DCN), sulfur-based species (H\(_2\)S, OC$^{33}$S, SO\(_2\)), and 12 COMs (including different isotopologues): CH\(_3\)OH, \(^{13}\)CH\(_3\)OH, CH\(_3\)\(^{18}\)OH, CH\(_2\)DOH, CH\(_3\)CHO, CH\(_3\)OCH\(_3\), CH\(_3\)OCHO, CH\(_3\)COCH\(_3\), C\(_2\)H\(_5\)OH, C\(_2\)H\(_5\)CN, \(^{13}\)CH\(_3\)CN, and CH\(_3\)C\(^{15}\)N.
The L1551~IRS~5~N spectrum includes 249 transitions from these 12 COMs.
The line emission from the Southern disk of L1551~IRS~5 is weaker, and we only found emission from a sub-set of the molecules found in the Northern disk (CH\(_3\)OH, CH\(_3\)OCH\(_3\), CH\(_3\)OCHO, \(^{13}\)CH\(_3\)CN, and H\(_2\)S) for a total of 37 COM transitions.
In the case of Haro~5a~IRS the continuum absorption is stronger than in the other targets decreasing the amount of species detected: CH\(_3\)OH, CH\(_3\)CHO, CH\(_3\)OCH\(_3\), CH\(_3\)OCHO, H\(_2\)S, and DCN.
The spectra of Haro~5a~IRS also include 46 transitions from the COMS, and show tentative emission from H\(_2\)CO and OC$^{33}$S but their low SNR prevent us from gaining any information about them.
The continuum in V346~Nor and OO~Ser did not show significant absorption.
However, comparing these two targets to L1551~IRS~5~N, we detected fewer COMs and fewer transitions in V346~Nor (6 COMs and 70 transitions) and OO~Ser (6 COMs and 51 transitions).
In V346~Nor we detected CH$_3$OH, CH$_2$DOH, CH$_3$CHO, CH$_3$OCH$_3$, CH$_3$OCHO, C$_2$H$_5$CN, H$_2$S, DCN, H$_2$CO, D$_2$CO, and SO$_2$.
Finally, in OO~Ser we detected CH$_3$OH, CH$_3$CHO, CH$_3$OCH$_3$, CH$_3$OCHO, CH$_3$COCH$_3$, H$_2$S, DCN, H$_2$CO, D$_2$CO, and SO$_2$.

\subsection{Spatial distribution of COMs}\label{ss:extension_coms}
\begin{figure*}
\centering
\includegraphics[width=\textwidth]{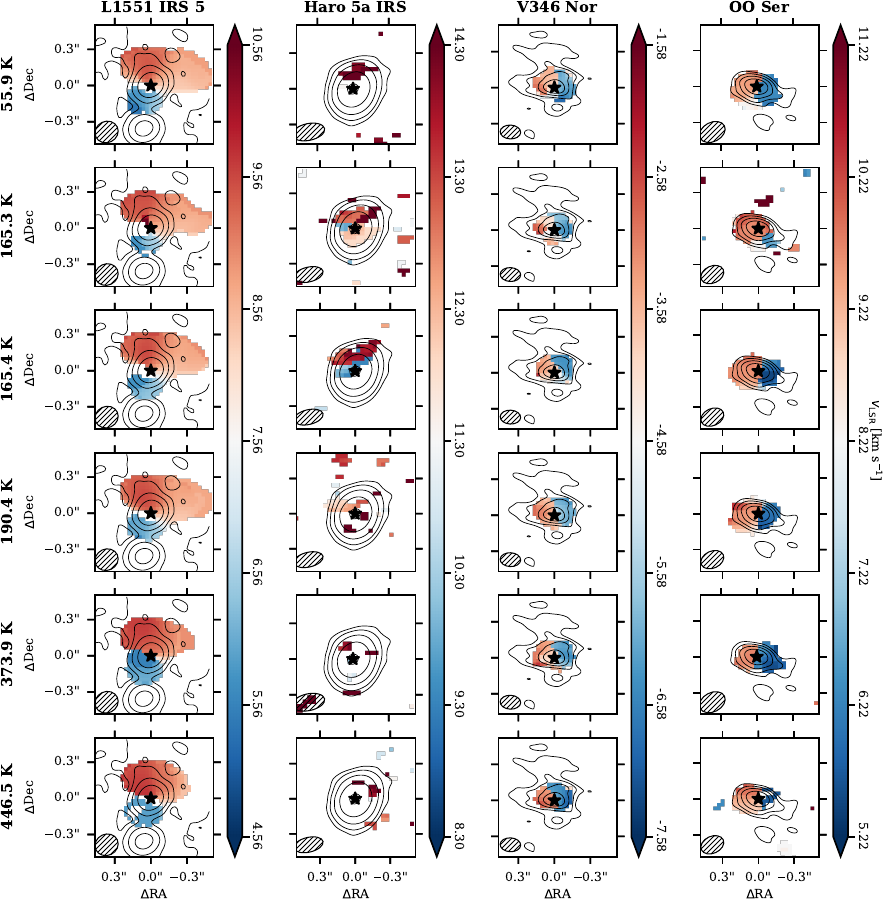}
\caption{Moment~1 maps of the different CH\(_3\)OH transitions found in our targets. From top to bottom in increasing $E_\mathrm{up}$. From left to right are L1551~IRS~5, Haro~5a~IRS, V346~Nor, and OO~Ser. The color-scale of each target have been centered on the systemic velocity of the protostar (see \autoref{tab:positions}), and are limited to $\pm$3\,km\,s$^{-1}$ with respect to this velocity. The contours trace the continuum emission and are at the same levels as in \autoref{fig:pixels}. The star symbols indicate the positions of the protostars. The beams are shown in the bottom left of each panel. The Southern disk of L1551~IRS~5 is not seen in the maps as it is quite faint and its systemic velocity is bluer than the limits of the color-scale of the plot. \label{fig:methanol_moments}}
\end{figure*}

In \autoref{fig:methanol_moments}, we present the Moment~1 maps of CH\(_3\)OH for each of our targets, which were created with a 3$\sigma$ cutoff, and in this subsection we discuss their morphologies.

Six transitions of CH\(_3\)OH are strongly detected in L1551~IRS~5~N, V346~Nor, and OO~Ser, and they show a velocity distribution that suggests disk rotation.
The Southern disk of L1551~IRS~5 is quite faint and its systemic velocity is bluer than the limits of the color-scale in the plot, thus this disk is not seen in the Moment maps.
Finally, the Haro~5a~IRS maps show few pixels with notable emission, which is due to the continuum absorption of its disk.
In the case of L1551~IRS~5, the transitions with lower $E_{up}$ values are more extended towards the east, which indicates a difference in the physical conditions of the material of the circumbinary disk or the envelope.
We suspect this extended emission traces material of the circumbinary disk or envelope that has been heated by the jets emanating from L1551~IRS~5 \citep{Fridlund1998_ApJ499L75F}.
However, the other three sources are also embedded and two of them (Haro~5a~IRS and V346~Nor) drive jets or outflows \citep{Kospal2017_ApJ84345K,CruzSaenzdeMiera2023ApJ94580C} yet we did not find a similar extended emission next to them.
The three disks have comparable inclinations \citep{Kospal2021_ApJS25630K}, which rules out a different perspective as an explanation for this lack of detection.
Instead, it can be a matter of lack of sensitivity and angular resolution because L1551~IRS~5 is 3 and 5 times closer to us than V346~Nor and Haro~5a~IRS, respectively.

We estimated the size of the CH\(_3\)OH emission from the disk by using the \textit{imfit} tool in CASA to fit a 2D Gaussian to the Moment~0 maps of the 232.78\,GHz transition of each target.
In the case of L1551~IRS~5~N, the transitions with $E_{up}<$~400\,K extend beyond the dust disk toward the West of the protostar, so we chose to only use the transition at 232.783\,GHz ($E_\mathrm{up}$ = 446.5\,K) to measure the extent of CH\(_3\)OH in all protostars.
Part of the output of \textit{imfit} is the semi-major full width at half maximum (FWHM), which we converted to a Gaussian $\sigma$ by dividing the FWHM by \(2\sqrt{2 \ln{2}}\), and we used this $\sigma$ as a proxy for the CH\(_3\)OH disk radius.

For L1551~IRS~5~North, the disk has a deconvolved major-axis FWHM of 0.521$\pm$0.108$''$, which gives a CH\(_3\)OH radius of 30\,au, an inclination of 46$\pm$17$^\circ$, and a position angle (PA) of 0.9$\pm$22.5$^\circ$.
This gas disk radius is a factor of 3 larger than the emission from the continuum at 1.3\,mm \citep{CruzSaenzdeMiera2019_ApJ882L4C}.
The CH\(_3\)OH emission from Southern disk was not resolved in most of the transitions, indicating that it has an extent of $<$12\,au, i.e., comparable to the extent of the disk in the continuum.
The continuum absorption of the Haro~5a~IRS disk caused that its CH\(_3\)OH maps only show a few pixels with significant emission, thus we were unable to compare its extent with respect to the dust.
For V346~Nor the best-fit values are a deconvolved major-axis FWHM of 0.177$\pm$0.017$''$ (resulting in a disk radius of 53$\pm$5\,au), an inclination of 24$\pm$17$^\circ$, and a PA of 96$\pm$82$^\circ$.
Therefore, the geometrical properties of the CH\(_3\)OH disk in V346~Nor are comparable, within the uncertainties, to the results obtained for the continuum emission via Gaussian fitting and via radiative transfer by \citet{Kospal2021_ApJS25630K}.
The disk of OO~Ser was barely resolved, and the fit resulted in a deconvolved major-axis FWHM of 0.155$\pm$0.068$''$, an inclination of 20$\pm$145$^\circ$, and a PA of 144$\pm$87$^\circ$.
The CH\(_3\)OH disk radius of OO~Ser is 29$\pm$12\,au which is the same as the dust radius found by \citet{Kospal2021_ApJS25630K}, and its inclination and PA are also in agreement with those from the dust disk.

\begin{figure*}
\centering
\includegraphics[height=0.9\textheight]{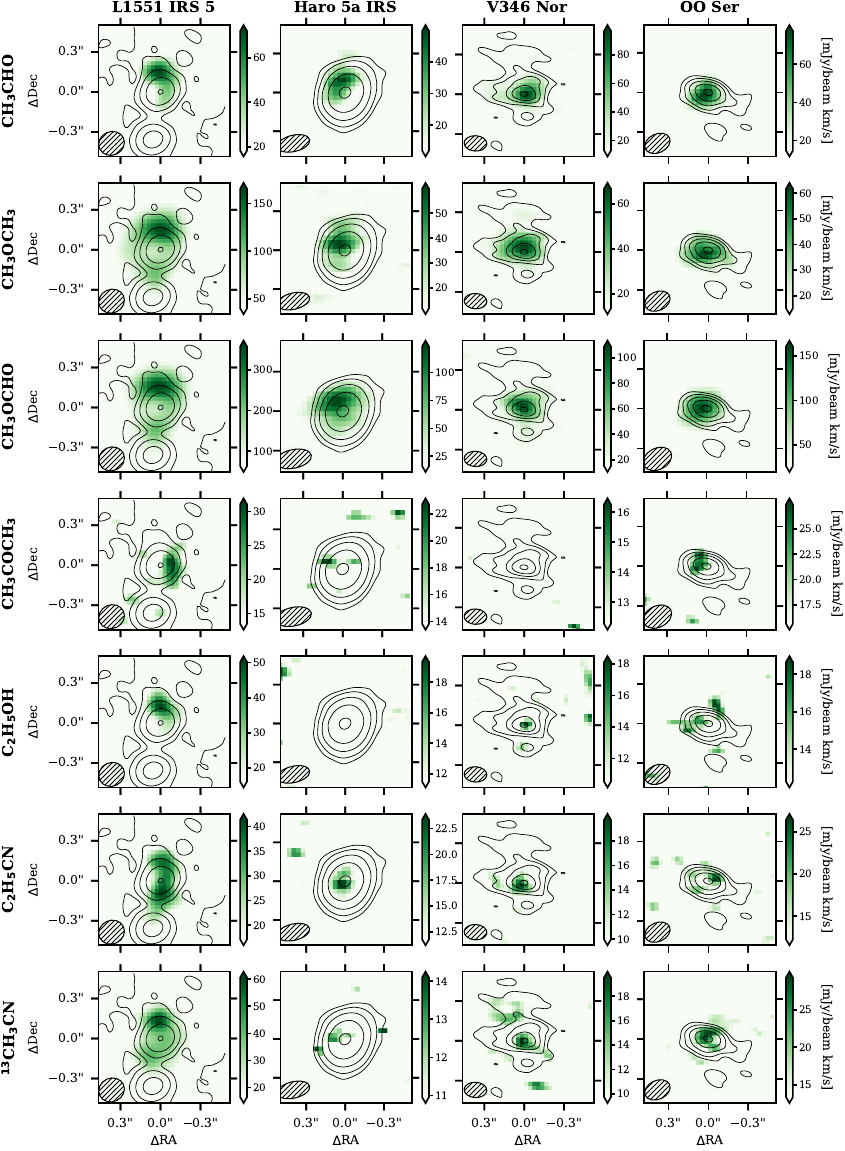}
\caption{Integrated emission maps of a representative line for 7 species on each of the FUor-type young stars. Each map has been centered around the peak of the continuum of the protostar. From top to bottom, the $E_\mathrm{up}$ values of the transitions are 108.3, 80.9, 111.5, 115.5, 175.3, 157.7, and 142.4\,K. The contours trace the continuum emission and are at the same levels as in \autoref{fig:pixels}.\label{fig:other_moments}}
\end{figure*}

In \autoref{fig:other_moments} we show the Moment~0 maps for different representative transitions of other COMs found in our targets.
In L1551~IRS~5, these COMs trace the emission of the Northern disk, while the Southern disk is only barely detected.
We measured the separations between the peaks in the Moment~0 maps of the COMs and the peaks of the continuum, and compared them to the semi-major axis of the beam of each map.
These separation are $\lesssim$25\% the size of the beam for both V346~Nor and OO~Ser, $\sim$30-40\% for Haro~5a~IRS, and $\sim$45-80\% for L1551~IRS~5~N.
We consider these separations to be an indication of the COM emission being off-center only for L1551~IRS~5~N and Haro~5a~IRS, and we corroborated this via a visual inspection.
In the case of Haro~5a~IRS, the emission peaks off-center, similar to CH\(_3\)OH, and its extent is much smaller than the dust disk.
For V346~Nor the emission is found close to the peak of the continuum, tracing an area similar to CH\(_3\)OH.
Finally, in the maps of OO~Ser, most of the COMs have emission close to the peak of the continuum.
Indeed, CH\(_3\)CHO, CH\(_3\)OCH\(_3\) and CH\(_3\)OCHO have morphologies similar to CH\(_3\)OH but CH\(_3\)COCH\(_3\) and C\(_2\)H\(_5\)CN peak off-center.

\begin{figure*}
\centering
\includegraphics[width=0.95\textwidth]{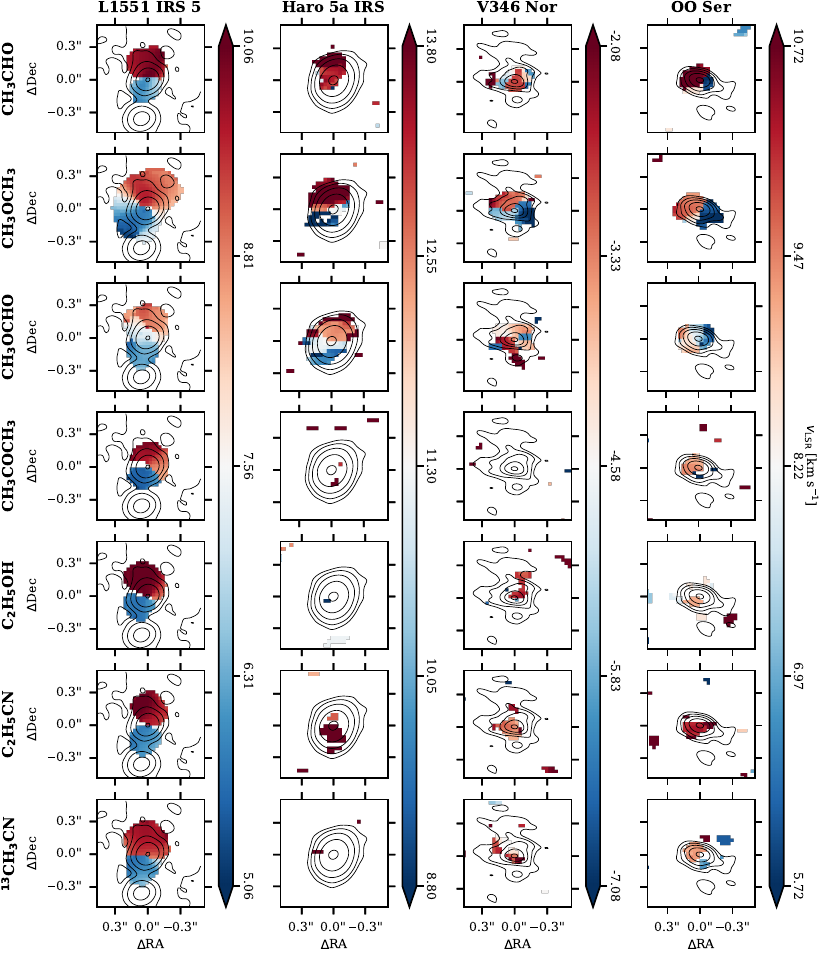}
\caption{Velocity field map (Moment 1) of the same representative transitions as \autoref{fig:other_moments}. The maps were generated with a 4$\sigma$ clipping to minimize the noise in the resulting image. The velocities are clipped at $\pm$3\,km\,s$^{-1}$. The velocities were centered on the $v_\mathrm{LSR}$ of each target (see \autoref{tab:positions}). The contours trace the continuum emission and are at the same levels as in \autoref{fig:pixels}.\label{fig:moment1}}
\end{figure*}

\autoref{fig:moment1} contains the velocity maps (moment 1) of the same representative transitions as \autoref{fig:other_moments}, which we use to examine which molecules in which targets show evidence of rotation.
The maps of L1551~IRS~5 show disk rotation for all the transitions in a southeast-northwest direction, comparable to the position angle of the disk found in the continuum \citep{CruzSaenzdeMiera2019_ApJ882L4C,Kospal2021_ApJS25630K}.
The Haro~5a~IRS disk only shows rotation in the maps of CH\(_3\)OCH\(_3\) and CH\(_3\)OCHO, which has an orientation also in agreement with the position angle estimated from the continuum \citep{Kospal2021_ApJS25630K}.
For the other detected COM, CH\(_3\)CHO, only the redshifted side is detected.
The remaining COMs were not detected and thus are not seen in the moment maps.
The redshifted emission seen in the C\(_2\)H\(_5\)CN map is from a nearby CH\(_3\)CHO transition.
In the case of V346~Nor, only CH\(_3\)OCH\(_3\) shows the velocity gradient expected from rotation.
For OO~Ser, CH\(_3\)CHO, CH\(_3\)OCH\(_3\) and CH\(_3\)OCHO indicate rotation.

\subsection{Spatial distribution of smaller molecules}\label{ss:extension_smaller}
\begin{figure*}[!ht]
\centering
\includegraphics[width=\textwidth]{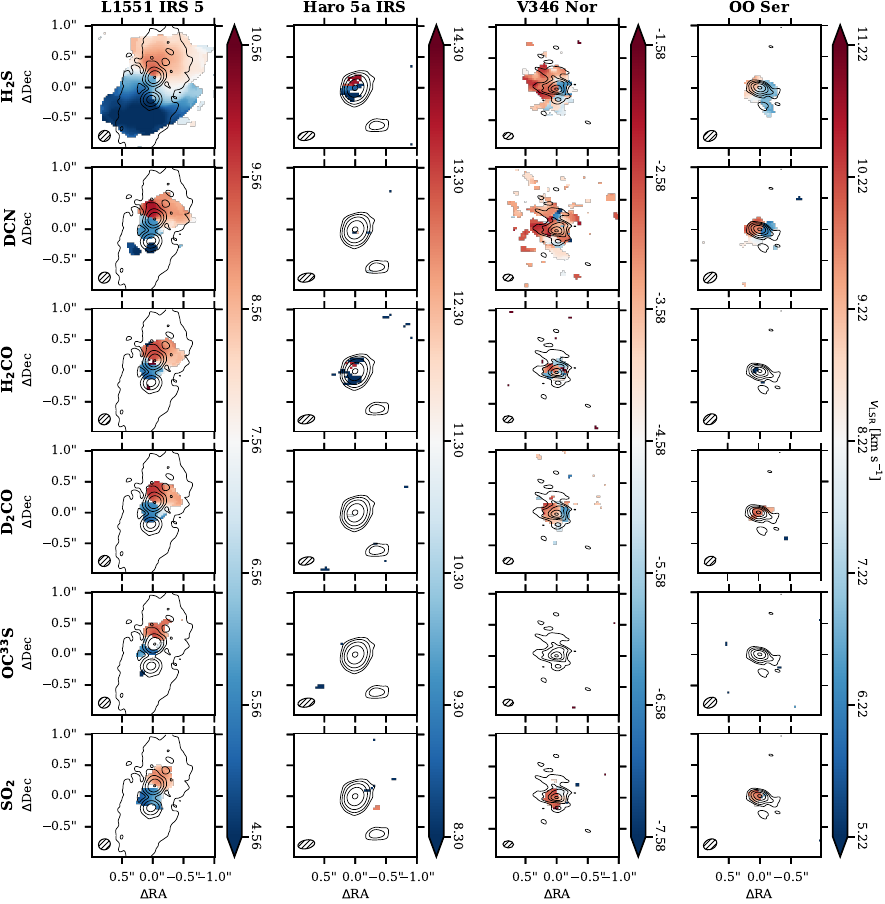}
\caption{Moment~1 maps of the small molecules found in our targets. From left to right are L1551~IRS~5, Haro~5a~IRS, OO~Ser, and V346~Nor. The color-scale of L1551~IRS~5 is centered around the average velocity of its two disks, while for the remainder of the targets the color-scales have been centered on the systemic velocity of each protostar. All the color-scales are limited to $\pm$3\,km\,s$^{-1}$. The contours trace the continuum emission and are at the same levels as in \autoref{fig:pixels}. The beams are shown in the bottom left of each panel.\label{fig:small_molecules}}
\end{figure*}

In \autoref{fig:small_molecules} we show the Moment~1 maps for the small molecules that we examined.
These maps were constructed with a 4$\sigma$ cutoff.
In this Figure, the color-scales for Haro~5a~IRS, V346~Nor, and OO~Ser were centered around their systemic velocities (see \autoref{tab:positions}).
For L1551~IRS~5, the color-scale was centered with respect to the average velocity of its two disks, which matches the systemic velocity obtained from APEX observations of $^{13}$CO $J$=3--2 \citep{CruzSaenzdeMiera2023ApJ94580C}.

As was the case for the COMs, the emission is stronger and most extended in L1551~IRS~5.
H\(_2\)S is the most extended and traces the circumbinary material.
This molecule shows high velocity emission in the outer parts of the protostellar disks (i.e.\ the northern side of the northern disk and the southern side of the southern disk) that aligns to where \citet{Takakuwa2020_ApJ89810T} suggested the circumbinary spiral arms make contact with both protostars.
The distributions of DCN, H\(_2\)CO, and D\(_2\)CO are similar to that of the lower \(E_{up}\) transitions of CH\(_3\)OH, including the extended emission toward the Western side of the Northern disk.
However, DCN also shows emission in the outer sides of both disks, at the same positions as in H\(_2\)S.
Meanwhile, the two remaining sulfur species, OC\(^{33}\)S and SO\(_2\), extend beyond the continuum emission.

Similar to the case of the COMs, the dust continuum in Haro~5a~IRS absorbs the emission from most of these small molecules, and only H\(_2\)S and H\(_2\)CO show emission with hints of disk rotation aligned with the dust disk \citep{Kospal2021_ApJS25630K}.

V346~Nor shows emission in all these small molecules except for OC\(^{33}\)S.
The small molecules extend beyond the COMs, and H\(_2\)S and DCN extend beyond the continuum emission, and H\(_2\)CO, D\(_2\)CO and SO\(_2\) have emission with an extent comparable to the continuum.
The distribution of velocities from these molecules is different from that of the COMs, which suggests contamination from the surrounded envelope.

Finally, the emission maps of OO~Ser show detections of H\(_2\)S, DCN, D\(_2\)CO, and SO\(_2\).
These four species have barely extended emission and their velocity profiles are consistent with rotation similar to the COMs.
H\(_2\)CO has emission in a few pixels close to the peak of the continuum.
H\(_2\)S in OO~Ser shows a blueshifted arm extending toward the south of the protostar, however, we lack the information to learn whether this is an embedded companion or signs of the surrounding envelope feeding material toward the FUor-like target similar to the recently reported streamer in \object{FU~Ori} \citep{Hales2024_ApJ96696H}.

\section{Analysis}\label{sec:analysis}
After identifying the detected molecules, we determined the column densities ($N$) and excitation temperatures ($T_\mathrm{ex}$) of the different species assuming local thermodynamic equilibrium (LTE), similar to what has been done for other hot corinos \citep[e.g.,][]{Jorgensen2016_AA595A117J,Coutens2016_AA590L6C,Coutens2018_AA612A107C}.
\citet{Andreu2023_AA677L17A} checked the LTE assumption for water isotopologues in L1551~IRS~5 and found consistent results using both LTE and non-LTE methods.
Thus, we assume that the other three targets have densities comparable to L1551~IRS~5 so that the LTE approximation is also valid for them.
We used the Markov chain Monte Carlo (MCMC) approach provided in CASSIS for our analysis.
This consists in generating synthetic spectra then comparing them to the observed data, and finding the set of parameters which results in the minimum $\chi^2$.
This MCMC approach enabled us to explore a wide parameter space.
Each synthetic spectrum is created using a combination of column density ($N$), excitation temperature ($T_\mathrm{ex}$), systemic velocity ($v_\mathrm{LSR}$), FWHM of the line, and source size ($s$).
A $\chi^2$ is computed for all the transitions selected for the comparison (meaning lines that are not blended and with sufficient SNR).
The uncertainties we consider for this $\chi^2$ are the rms of each spectral window (see \autoref{tab:observations}) and the nominal flux calibration uncertainty of ALMA \citep[10\%; ][]{ALMA_Cycle4}.
Using this methodology instead of a rotational diagram approach allows us to verify that all the transitions in the full spectral window (not just those used in the fit) match the observational data.
In addition, this approach of comparing synthetic spectra to observations does not assume that all the lines are optically thin, and instead the use of moderately optically thick lines can help constrain the source size of the analyzed species.

\href{https://zenodo.org/records/14967612}{Table A.1} lists all the identified transitions
We indicated the lines that are detected in each of the sources with diamond symbols.
A filled diamond means that the transition is used in the fitting procedure, and the transitions with open markers are ignored due to either blending or very high optical depths.

\subsection{L1551~IRS~5~North}
Our first MCMC runs for this disk consisted of letting the column density, excitation temperature, the systemic velocity, the FWHM of the line, and source size be free parameters for each species. 
The parameter space that we explored with MCMC for all targets was
10$^{12}$~cm$^{-2}\leq N \leq 10^{19}$~cm$^{-2}$,
10\,K~$\leq T_\mathrm{ex} \leq$~300~K,
\kms{7.0}$\leq v_\mathrm{LSR} \leq$ \kms{12.5},
\kms{1}$\leq \mathrm{FWHM} \leq$ \kms{10}, and
{0.01$''$}$\leq s \leq$ {1.0$''$}.
$N$, $T_\mathrm{ex}$ and $s$ were explored in logarithmic scale, while linear scales were used for $v_\mathrm{LSR}$ and $\mathrm{FWHM}$.
We ran the MCMC for 50\,000 iterations and dropped the first 45\,000 as a burn-in phase.
Afterwards we determined the best-fit parameters and their uncertainties using the 0.16, 0.50, and 0.84 percentiles.
We found that shortening the burn-in phase did not cause changes in the values of the best-fit parameters.

In our fitting procedure for CH\(_2\)DOH, we followed the recommendations provided by the JPL database and only included the lines with $K_a$~$<$~10 in the upper transition.
Furthermore, based on the analysis by \citet{Oyama2023_ApJ9574O}, we only kept the CH\(_2\)DOH transitions that were present in the JPL database and in their list, and those with $J$~$<$~26.

After obtaining a first version of the best-fit spectrum for all species, we incorporated it into a ``reference'' spectrum, which was used to double-check that the lines used in the fitting process were not blended.
The full list of best-fit values and their statistical uncertainties for this first set of MCMC runs can be found in \autoref{tab:north_bestfit}.
We emphasize that these uncertainties are purely statistical and should be considered as underestimated.
The subtraction of the baseline is known to be challenging for sources with very rich line spectra.
Among the sources studied here, L1551~IRS~5~North is the one with the highest number of lines.
We consequently used that source to estimate how an error on the baseline subtraction could affect the column densities.
We assumed that our continuum level was overestimated by 1 \textit{rms} so we added 1 \textit{rms} to the spectrum and performed a new set of fits.
In this additional fitting process, we fixed $s$, $v_\mathrm{LSR}$, and $T_\mathrm{ex}$ to the values obtained when all parameters were initially free (i.e., top panel of \autoref{tab:north_bestfit}).
We then determined the best-fit column densities and FWHM, and found that the values of $N$ were higher by a factor between 1.2 to 1.8, depending on the COM.
As the spectra of the other sources are less line crowded, this factor is expected to be an upper limit for the other sources.

\begin{table*}
  \caption{Best-fit parameters obtained for the L1551~IRS~5 Northern disk using two different approaches.\label{tab:north_bestfit}}
  \centering
    \begin{tabular}[c]{ccccd{4.3}@{\hskip 0.4in}cd{2.6}c}
      \hline\\[-9pt]
      Species & n$_{\mathrm{lines}}$ & $E_\mathrm{up}$ [K] & $N$ [cm$^{-2}$] & \multicolumn{1}{c}{$T_{\mathrm{ex}}$ [K]} & FWHM [km\,s$^{-1}$] & \multicolumn{1}{c}{$v_\mathrm{LSR}$ [km\,s$^{-1}$]}& $s$ [$''$]\\
      \hline
      \hline\\[-9pt]
\multicolumn{8}{c}{COMs with free parameters}\\
\hline\\[-9pt]
CH$_3$OH          & 18 & 56 -- 946  & \uncsci{1.72}{-0.08}{+0.08}{19} & \unc{294.90}{-4.40}{+3.41}   & \unc{1.97}{-0.03}{+0.03} & \unc{9.59}{-0.01}{+0.02}  & \unc{0.05}{-0.01}{+0.01}\\
CH$_3$OH$^a$      & 7  & 657 -- 946 & \uncsci{1.11}{-0.12}{+0.17}{20} & \unc{140.30}{-4.60}{+3.90}   & \unc{1.92}{-0.04}{+0.06} & \unc{9.77}{-0.03}{+0.03}  & \unc{0.05}{-0.01}{+0.01}\\
$^{13}$CH$_3$OH   & 3  & 162 -- 254 & \uncsci{1.22}{-0.10}{+0.25}{17} & \unc{128.40}{-13.00}{+9.20}  & \unc{1.98}{-0.09}{+0.12} & \unc{9.72}{-0.04}{+0.04}  & \unc{0.25}{-0.06}{+0.03}\\
CH$_3${$^{18}$O}H & 9  & 33 -- 96   & \uncsci{1.12}{-0.10}{+0.16}{16} & \unc{91.31}{-8.76}{+7.98}    & \unc{2.56}{-0.16}{+0.21} & \unc{9.75}{-0.05}{+0.05}  & \unc{0.42}{-0.04}{+0.04}\\
CH$_2$DOH         & 9  & 49 -- 352  & \uncsci{6.57}{-0.54}{+0.57}{18} & \unc{145.60}{-5.30}{+4.90}   & \unc{1.94}{-0.13}{+0.13} & \unc{10.06}{-0.06}{+0.06} & \unc{0.05}{-0.01}{+0.01}\\
CH$_2$DOH$^b$     & 6  & 49 -- 352  & \uncsci{5.25}{-0.72}{+0.90}{18} & \unc{210.30}{-12.50}{+24.70} & \unc{1.78}{-0.04}{+0.05} & \unc{9.98}{-0.02}{+0.03} & \unc{0.06}{-0.01}{+0.01}\\
CH$_3$CHO         & 27 & 65 -- 387  & \uncsci{1.12}{-0.15}{+0.23}{17} & \unc{212.30}{-6.30}{+8.40}   & \unc{2.28}{-0.05}{+0.06} & \unc{9.80}{-0.02}{+0.02}  & \unc{0.07}{-0.01}{+0.01}\\
CH$_3$OCH$_3$     & 8  & 81 -- 253  & \uncsci{9.57}{-0.51}{+0.65}{16} & \unc{184.60}{-5.90}{+6.20}   & \unc{2.59}{-0.06}{+0.06} & \unc{9.31}{-0.03}{+0.03}  & \unc{0.26}{-0.01}{+0.01}\\
CH$_3$OCHO        & 34 & 106 -- 412 & \uncsci{3.58}{-0.22}{+0.30}{17} & \unc{162.50}{-8.48}{+5.40}   & \unc{2.11}{-0.04}{+0.05} & \unc{9.55}{-0.02}{+0.02}  & \unc{0.09}{-0.01}{+0.01}\\
CH$_3$COCH$_3$    & 13 & 111 -- 116 & \uncsci{6.55}{-3.35}{+5.46}{16} & \unc{104.60}{-10.40}{+10.10} & \unc{2.38}{-0.22}{+0.26} & \unc{9.60}{-0.08}{+0.07}  & \unc{0.05}{-0.01}{+0.02}\\
C$_2$H$_5$OH      & 24 & 24 -- 464  & \uncsci{5.24}{-2.04}{+1.52}{17} & \unc{190.70}{-23.30}{+19.60} & \unc{2.23}{-0.10}{+0.09} & \unc{9.95}{-0.03}{+0.03}  & \unc{0.07}{-0.01}{+0.02}\\
C$_2$H$_5$CN      & 20 & 147 -- 368 & \uncsci{2.43}{-0.07}{+0.07}{15} & \unc{190.60}{-5.20}{+5.30}   & \unc{2.48}{-0.08}{+0.07} & \unc{9.74}{-0.03}{+0.03}  & \unc{0.42}{-0.02}{+0.02}\\
$^{13}$CH$_3$CN   & 5  & 78 -- 257  & \uncsci{4.71}{-1.30}{+1.73}{15} & \unc{170.60}{-18.39}{+15.50} & \unc{2.11}{-0.09}{+0.11} & \unc{9.62}{-0.04}{+0.03}  & \unc{0.06}{-0.01}{+0.01}\\
CH$_3$C{$^{15}$N} & 3  & 78 -- 335  & \uncsci{4.64}{-0.78}{+0.95}{14} & \unc{242.55}{-35.53}{+37.25} & \unc{2.77}{-0.38}{+0.83} & \unc{9.27}{-0.18}{+0.17}  & \unc{0.43}{-0.12}{+0.04}\\
      \hline
      \hline\\[-9pt]
\multicolumn{8}{c}{COMs with $T_\mathrm{ex}$~=~177.6\,K, FWHM~=~2.17\,km\,s$^{-1}$, $v_\mathrm{LSR}$~=~9.67\,km\,s$^{-1}$ and $s$~=~0.08$''$}\\
      \hline
CH$_3$OH          & 18 & 56 -- 946  & \uncsci{4.66}{-0.18}{+0.16}{18} &  &  &  & \\
CH$_3$OH$^a$      & 7  & 657 -- 946 & \uncsci{1.60}{-0.07}{+0.06}{19} &  &  &  & \\
$^{13}$CH$_3$OH   & 3  & 162 -- 254 & \uncsci{4.97}{-0.27}{+0.30}{17} &  &  &  & \\
CH$_3${$^{18}$O}H & 9  & 33 -- 96   & \uncsci{9.94}{-0.49}{+0.49}{16} &  &  &  & \\
CH$_2$DOH         & 9  & 49 -- 352  & \uncsci{1.08}{-0.06}{+0.07}{18} &  &  &  & \\
CH$_2$DOH$^b$     & 6  & 49 -- 352  & \uncsci{2.81}{-0.17}{+0.19}{18} &  &  &  & \\
CH$_3$CHO         & 27 & 65 -- 387  & \uncsci{8.04}{-0.17}{+0.18}{16} &  &  &  & \\
CH$_3$OCH$_3$     & 8  & 81 -- 253  & \uncsci{3.41}{-0.10}{+0.11}{17} &  &  &  & \\
CH$_3$OCHO        & 34 & 106 -- 412 & \uncsci{5.05}{-0.11}{+0.11}{17} &  &  &  & \\
CH$_3$COCH$_3$    & 13 & 111 -- 116 & \uncsci{1.10}{-0.07}{+0.07}{17} &  &  &  & \\
C$_2$H$_5$OH      & 24 & 24 -- 464  & \uncsci{3.39}{-0.12}{+0.12}{17} &  &  &  & \\
C$_2$H$_5$CN      & 20 & 147 -- 368 & \uncsci{8.84}{-0.23}{+0.22}{15} &  &  &  & \\
$^{13}$CH$_3$CN   & 5  & 78 -- 257  & \uncsci{2.73}{-0.12}{+0.11}{15} &  &  &  & \\
CH$_3$C{$^{15}$N} & 3  & 78 -- 335  & \uncsci{1.15}{-0.15}{+0.16}{15} &  &  &  & \\
      \hline
    \end{tabular}
    \tablefoot{
 The top part of the table are the results from the runs with all parameters begin left free, and the bottom part when fixing the $T_\mathrm{ex}$, FWHM, $v_\mathrm{LSR}$ and $s$ to the median values obtained from the all-free runs. The uncertainties listed in each column only represent the statistical uncertainties from the fitting procedure. For CH\(_3\)OH we include the results obtained using only the high excitation lines (labeled with $^a$), and for CH$_2$DOH we include the results obtained using only the lines with the lowest opacities (labeled with $^b$). The second and third columns are the amount of transitions used in the fit and their $E_\mathrm{up}$ range.
    }
\end{table*}

The observed spectrum of CH\(_3\)OH in L1551~IRS~5~North includes transitions with 56\,K~$\lesssim$~$E_\mathrm{up}$~$\lesssim$~946\,K.
The best-fit model is dominated by the lower-excitation lines with $E_\mathrm{up}$~$\lesssim$~335\,K (see \href{https://zenodo.org/records/14967612}{Figure B.1}) and the fluxes of the high excitation lines are underestimated. 
As the low-excitation lines are commonly highly optically thick in hot corinos \citep[e.g.,][]{Jorgensen2016_AA595A117J,Bianchi2020_MNRAS498L87B}, we ran our fitting again using only lines with $E_\mathrm{up}$ $\gtrsim$ 657\,K. 
After comparing the results from both fits, we found that $s$, FWHM and the $v_\mathrm{LSR}$ did not change significantly.  
However, the best-fit values of $N$ and $T_\mathrm{ex}$ are different.
This high-excitation fitting results in a higher $N$ and lower $T_\mathrm{ex}$ than the one including all the CH\(_3\)OH lines (see \autoref{tab:north_bestfit}).
Using highly optically thick lines could lead to an underestimation of the column density.
So by using less optically thick high-excitation lines only, our best-fit $N$ should be more accurate.
To test this, we computed the column densities of CH\(_3\)OH using its less abundant isotopologues,\(^{13}\)CH\(_3\)OH and CH\(_3\)\(^{18}\)OH, and assuming standard isotopic ratios from the literature: [$^{12}$C]/[$^{13}$C]~=~68 \citep{Milam2005_ApJ6341126M} and [$^{16}$O]/[$^{18}$O]~=~560 \citep{WilsonRood1994_ARA&A32191W}.
For this test, we fixed the $T_\mathrm{ex}$, FWHM, $v_\mathrm{LSR}$ and $s$ to the median values of the best-fits found for all COMs (assuming all the CH$_3$OH isotopologues and the COMs come from the same region), which left the column density as the only free parameter.
When determining the medians, we excluded the results of CH$_3$OH (and CH$_2$DOH) that were obtained using optically thick lines.
Here we ran the MCMC for 10\,000 iterations and  dropped the first 5\,000 as a burn-in phase.
We multiplied the best-fit $N$ by the aforementioned isotopic ratios and computed methanol column density values of \sci{3.38}{19}\,cm$^{-2}$ and \sci{5.57}{19}\,cm$^{-2}$ for the $^{13}$C and $^{18}$O isotopologues, respectively.
These are in reasonable agreement with the fit of the high-excitation lines (\sci{1.60}{19}\,cm$^{-2}$) by factors of 2--3, and are an order of magnitude higher than the value obtained when using the optically thick lines (\sci{4.66}{18}\,cm$^{-2}$).
Therefore, for the remaining of the paper, we will only utilize the column density obtained using the high-excitation lines.
The result of this high-excitation fit is marked with an $^a$ in \autoref{tab:north_bestfit}.

A similar problem is seen for CH$_2$DOH.
The high opacity lines are well fitted while the fluxes of the optically thin lines are underestimated.
So we ran the MCMC again only including the transitions with the lowest opacities.
The fit obtained when including the optically thick lines resulted in a higher $T_\mathrm{ex}$ (210\,K versus 146\,K) and a lower $N$ (\sci{5.25}{18}\,cm$^{-2}$ versus \sci{6.57}{18}\,cm$^{-2}$).
The results are included in \autoref{tab:north_bestfit}, labeled with a $^b$.
Just like for methanol, for the remainder of the paper, we will work with the fit found using only the low opacity lines.

For the other COMs, we also ran the second batch of MCMC runs where we fixed the $T_\mathrm{ex}$, $v_\mathrm{LSR}$, FWHM and $s$ parameters to the medians of the best-fit values previously determined assuming free parameters. 
The models are of comparable quality with respect to the observational data as the ones obtained using free parameters (see \href{https://zenodo.org/records/14967612}{Appendix B}).
For the majority of the lines, the differences between the data (black line) and the total model (green line) are below the 3$\sigma$ level (shaded area).
The full list of best-fit values and their statistical uncertainties for both batches of MCMC runs can be found in \autoref{tab:north_bestfit}.

\subsection{L1551~IRS~5~South}
The Southern disk of L1551~IRS~5 has fewer and weaker lines than its Northern counterpart.
Due to a lack of constraints, we assumed for these MCMC runs the same source size as the Northern disk (0.08$''$). 
The first set of MCMC runs were with $N$, $T_\mathrm{ex}$, $v_\mathrm{LSR}$ and FWHM left as free parameters.
The parameter space explored for these variables was the same as for the Northern disk, except for $v_\mathrm{LSR}$ which we limited to within 2 and 5\,km\,s$^{-1}$.

We computed the medians of this first set of MCMC runs and found $T_\mathrm{ex}$ = 193.90\,K, FWHM = 2.23\,km\,s$^{-1}$ and $v_\mathrm{LSR}$ = 3.47\,km\,s$^{-1}$.
For the computation of the medians, we excluded the results for CH$_3$OCH$_3$ because its best-fit $T_\mathrm{ex}$ value cannot be constrained due to the four lines detected in this species having the same E$_{up}$ value.
The median values of $T_\mathrm{ex}$ and FWHM are the same as those found independently for CH$_3$OH.
However, the best-fit value of $v_\mathrm{LSR}$ found for CH\(_3\)OH is \kms{0.66} bluer that the median $v_\mathrm{LSR}$, which is large enough to cause a bad fit of CH\(_3\)OH.
It is unclear whether the discrepancy between $v_\mathrm{LSR}$ values is real or an effect caused by absorption at redshifted velocities. 
We carried out a second set of MCMC runs for CH$_3$OCH$_3$, CH$_3$OCHO and $^{13}$CH$_3$CN by fixing $T_\mathrm{ex}$, $v_\mathrm{LSR}$ and FWHM to the medians found from the first run.
This second set of column densities will be the ones used to compute the abundances w.r.t.\ methanol (see below) even though CH$_3$OH has a different $v_\mathrm{LSR}$ than the other three COMs.
These abundance ratios could then be quite uncertain.

The best-fit values and their statistical uncertainties for both MCMC runs, and the median values used for the second set of runs, are presented in \autoref{tab:south_bestfit}.
Similar to its Northern counterpart, the models in L1551~IRS~5~S (\href{https://zenodo.org/records/14967612}{Appendix C}) match the data, with the uncertainties being less than the 3$\sigma$ level.
\begin{table*}
  \caption{Similar to \autoref{tab:north_bestfit} but for the Southern disk of L1551~IRS~5 assuming a source size of 0.08$''$.\label{tab:south_bestfit}}
  \begin{center}
    \begin{tabular}[c]{ccccd{4.3}@{\hskip 0.4in}cd{2.6}}
      \hline\\[-9pt]
      Species & n$_{\mathrm{lines}}$ & $E_\mathrm{up}$ [K] & $N$ [cm$^{-2}$] & \multicolumn{1}{c}{$T_{\mathrm{ex}}$ [K]} & FWHM [km\,s$^{-1}$] & \multicolumn{1}{c}{$v_\mathrm{LSR}$ [km\,s$^{-1}$]}\\
      \hline
      \hline\\[-9pt]
\multicolumn{7}{c}{COMs with free parameters}\\
      \hline\\[-9pt]
CH$_3$OH        & 8  & 56 -- 746  & \uncsci{3.99}{-0.24}{+0.24}{17} & \unc{193.90}{-11.20}{+10.90} & \unc{2.23}{-0.09}{+0.12} & \unc{2.81}{-0.04}{+0.05}\\
CH$_3$OCH$_3$   & 4  & 81         & \uncsci{2.92}{-0.26}{+0.35}{16} & \unc{88.80}{-7.14}{+10.30}   & \unc{2.65}{-0.21}{+0.21} & \unc{3.45}{-0.07}{+0.06}\\
CH$_3$OCHO      & 15 & 106 -- 312 & \uncsci{5.00}{-0.62}{+0.57}{16} & \unc{200.25}{-12.15}{+12.27} & \unc{1.79}{-0.18}{+0.18} & \unc{3.52}{-0.11}{+0.10}\\
$^{13}$CH$_3$CN & 4  & 78 -- 142  & \uncsci{4.30}{-0.52}{+0.69}{14} & \unc{83.25}{-12.65}{+29.88}  & \unc{2.96}{-0.31}{+0.27} & \unc{3.49}{-0.17}{+0.16}\\
      \hline
      \hline\\[-9pt]
\multicolumn{7}{c}{COMs with $T_\mathrm{ex}$~=~193.9\,K, FWHM~=~2.23\,km\,s$^{-1}$ and $v_\mathrm{LSR}$~=~3.47\,km\,s$^{-1}$}\\
      \hline\\[-9pt]
CH$_3$OCH$_3$   & 4  & 81         & \uncsci{6.86}{-0.36}{+0.39}{16} & & & \\
CH$_3$OCHO      & 15 & 106 -- 312 & \uncsci{5.58}{-0.21}{+0.21}{16} & & & \\
$^{13}$CH$_3$CN & 4  & 78 -- 142  & \uncsci{5.87}{-0.61}{+0.61}{14} & & & \\
      \hline
    \end{tabular}
  \tablefoot{In the case of a single value of $E_\mathrm{up}$ when n$_\mathrm{lines}>1$, it means that those lines are all within 1\,K of each other.}
  \end{center}
\end{table*}

\subsection{Haro~5a~IRS}
In the case of Haro~5a~IRS, we were not able to constrain the source size.
There is a degeneracy between the column densities and the source size.
Consequently, we decided to set the source size to 0.1$''$, which is comparable to the size of the emission seen in the moment maps (see \autoref{fig:methanol_moments} and \autoref{fig:other_moments}).
We also had to fix the FWHM at 2.5\,km\,s$^{-1}$ in order to decrease the number of free parameters and better fit the low SNR lines.
This FWHM value was measured with CASSIS by fitting a Gaussian profile on the lines of the different COMs separately.
The parameter space of $N$ and $T_\mathrm{ex}$ were the same as for the two previous disks, and for the $v_\mathrm{LSR}$, it was between 13 and 15\,km\,s$^{-1}$.

As for the two previous disks, a second set of MCMC runs were done using the median values of $T_\mathrm{ex}$ and $v_\mathrm{LSR}$.
However, after a visual inspection of the quality of these follow-up fits, we found that the models were not a good fit for all the molecules.
We ran a third set of MCMC fits using the $T_\mathrm{ex}$ and $v_\mathrm{LSR}$ from the CH\(_3\)OH fit, and found that the resulting models were a slightly better fit than the ones using the medians.
Thus, we kept the models obtained using the CH\(_3\)OH constraints.
We examined the residuals between the data and the best-fit model and found that the residual of all lines used for the fit fall within $\pm$3$\sigma$.
The best-fit values and their statistical uncertainties are in \autoref{tab:haro5airs_bestfit}, and the best-fit spectra are shown in \href{https://zenodo.org/records/14967612}{Appendix D}.

\begin{table*}
  \caption{Best-fit parameters obtained for Haro~5a~IRS assuming a source size of 0.1$''$.\label{tab:haro5airs_bestfit}}
  \begin{center}
    \begin{tabular}[c]{ccccd{4.3}@{\hskip 0.4in}cd{2.6}}
      \hline\\[-9pt]
      Species & n$_{\mathrm{lines}}$ & $E_\mathrm{up}$ [K] & $N$ [cm$^{-2}$] & \multicolumn{1}{c}{$T_{\mathrm{ex}}$ [K]} & FWHM [km\,s$^{-1}$] & \multicolumn{1}{c}{$v_\mathrm{LSR}$ [km\,s$^{-1}$]}\\
      \hline
      \hline\\[-9pt]
\multicolumn{7}{c}{COMs with free parameters and FWHM~=~2.50\,km\,s$^{-1}$}\\
      \hline\\[-9pt]
CH$_3$OH      & 5  & 56 -- 374  & \uncsci{9.88}{-1.13}{+1.19}{16} & \unc{225.10}{-23.77}{+21.90} &  & \unc{13.76}{-0.23}{+0.25} \\
CH$_3$CHO     & 13 & 65 -- 129  & \uncsci{6.36}{-0.41}{+0.53}{15} & \unc{110.60}{-7.40}{+6.64}   &  & \unc{13.42}{-0.06}{+0.07} \\
CH$_3$OCH$_3$ & 4  & 81         & \uncsci{2.30}{-0.21}{+0.48}{16} & \unc{127.50}{-7.70}{+24.60}  &  & \unc{13.71}{-0.04}{+0.10} \\
CH$_3$OCHO    & 3  & 109 -- 109 & \uncsci{1.67}{-0.20}{+0.60}{16} & \unc{87.50}{-18.70}{+28.75}  &  & \unc{13.72}{-0.12}{+0.11} \\
      \hline
      \hline\\[-9pt]
\multicolumn{7}{c}{COMs with $T_\mathrm{ex}$~=~225.10\,K, FWHM~=~2.50\,km\,s$^{-1}$ and $v_\mathrm{LSR}$~=~13.76\,km\,s$^{-1}$}\\
      \hline\\[-9pt]
CH$_3$OH      & 5  & 56 -- 374  & \uncsci{9.89}{-0.70}{+0.64}{16} &  &  & \\
CH$_3$CHO     & 13 & 65 -- 129  & \uncsci{1.34}{-0.08}{+0.07}{16} &  &  & \\
CH$_3$OCH$_3$ & 4  & 81         & \uncsci{5.54}{-0.30}{+0.29}{16} &  &  & \\
CH$_3$OCHO    & 3  & 109 -- 109 & \uncsci{3.71}{-0.27}{+0.28}{16} &  &  & \\
      \hline
    \end{tabular}
  \tablefoot{In the case of a single value of $E_\mathrm{up}$ when n$_\mathrm{lines}>1$, it means that those lines are all within 1\,K of each other.}
  \end{center}
\end{table*}

\subsection{V346~Nor}
For V346~Nor, the source size was also not constrained.
So we assumed the size determined from the integrated emission map of CH\(_3\)OH, 0.177$''$ (see \autoref{ss:extension_coms}). 
We first ran the MCMC with $N$, $T_\mathrm{ex}$, $v_\mathrm{LSR}$ and FWHM as free parameters.
The parameter space for $v_\mathrm{LSR}$ was from -5.0\,km\,s$^{-1}$ to -2.0\,km\,s$^{-1}$ while for the other parameters it was the same as for the other disks.
The CH\(_3\)CHO lines appear to experience absorption at redshifted velocities, causing the best-fit $v_\mathrm{LSR}$ value to be \kms{0.3} lower compared to the other COMs.
The best-fit values and their statistical uncertainties are in \autoref{tab:v346nor_bestfit}, and the best-fit spectra are shown in \href{https://zenodo.org/records/14967612}{Appendix E}.

After fitting the COMs with all free parameters, we ran a secondary fitting using the median values obtained from it for the second set of runs.
Just as the case of Haro~5a~IRS, the models obtained using the median values was not satisfactory, and we ran a third set of fits using the results of the CH\(_3\)OH fit.
We carried out a visual comparison between the models obtained using the medians and the models obtained with the values found for CH\(_3\)OH.
In the case of CH\(_3\)OH, we find that the model computed using the medians underestimates the fluxes of three lines by factors between 15\% and 50\%.
For CH\(_3\)CHO, the medians model results in six transitions with fluxes $\leq$25\% higher than the observed spectrum, whereas the lines from the CH\(_3\)OH model fit the data in shape and scale.
Thus, the CH\(_3\)OH model is better at describing CH\(_3\)CHO.
In CH\(_3\)OCH\(_3\) and CH\(_3\)OCHO, both models represent these molecules equally well.
For C\(_2\)H\(_5\)CN, the CH\(_3\)OH model slightly underestimates the flux of one line by $\sim$25\%, while the medians model overestimates one line by $\sim$10\%.
Due to the higher SNR of the CH\(_3\)OH and CH\(_3\)CHO lines, we decided to keep the column densities obtained using the CH\(_3\)OH parameters for the remaining of this paper.

We noted that 18 lines in this source have line profiles different from the synthetic ones (see \href{https://zenodo.org/records/14967612}{Appendix E}).
Indeed, the majority of the CH\(_3\)OH lines in V346~Nor (\href{https://zenodo.org/records/14967612}{Figure E.1}) have flatter line profiles.
For the lines with strong emission (panels $a$, $b$, $c$, $d$, $f$, $g$, and $j$) it could indicate that they are slightly optically thick (which suggests that our column density is a lower limit, as discussed below).
For the other weaker lines, they are too close to the noise level to accurately discuss their line profiles.
Some CH\(_3\)CHO lines also differ from the synthetic spectrum, featuring a narrow blueshifted peak (see panels $a$, $b$, $f$, $j$, and $m$ of \href{https://zenodo.org/records/14967612}{Figure E.4}).
However, the angular and spectral resolutions of the observations are not sufficient to identify what causes these different line profiles.

\begin{table*}
  \caption{Best-fit parameters obtained for V346~Nor assuming a source size of 0.177$''$.\label{tab:v346nor_bestfit}}
  \centering
    \begin{tabular}[c]{ccccd{4.3}@{\hskip 0.4in}cd{2.6}}
      \hline\\[-9pt]
      Species & n$_{\mathrm{lines}}$ & $E_\mathrm{up}$ [K] & $N$ [cm$^{-2}$] & \multicolumn{1}{c}{$T_{\mathrm{ex}}$ [K]} & FWHM [km\,s$^{-1}$] & \multicolumn{1}{c}{$v_\mathrm{LSR}$ [km\,s$^{-1}$]}\\
      \hline
      \hline\\[-9pt]
\multicolumn{7}{c}{COMs with free parameters}\\
      \hline\\[-9pt]
CH$_3$OH      & 13 & 56 -- 835  & \uncsci{2.06}{-0.04}{+0.04}{18} & \unc{210.70}{-3.00}{+3.90}   & \unc{5.43}{-0.06}{+0.10} & \unc{-4.66}{-0.01}{+0.05}\\
CH$_2$DOH     & 3  & 113 -- 167 & \uncsci{5.98}{-1.34}{+0.92}{17} & \unc{245.40}{-53.10}{+36.37} & \unc{3.91}{-0.16}{+0.16} & \unc{-4.71}{-0.09}{+0.08}\\
CH$_3$CHO     & 17 & 65 -- 357  & \uncsci{1.71}{-0.14}{+0.15}{16} & \unc{123.30}{-12.70}{+13.30} & \unc{5.04}{-0.18}{+0.24} & \unc{-5.46}{-0.10}{+0.11}\\
CH$_3$OCH$_3$ & 8  & 81 -- 253  & \uncsci{7.28}{-0.67}{+0.73}{16} & \unc{152.00}{-11.30}{+9.30}  & \unc{5.37}{-0.15}{+0.11} & \unc{-4.73}{-0.04}{+0.03}\\
CH$_3$OCHO    & 7  & 106 -- 312 & \uncsci{6.12}{-0.35}{+0.49}{16} & \unc{156.40}{-8.10}{+11.80}  & \unc{4.99}{-0.33}{+0.18} & \unc{-4.69}{-0.13}{+0.08}\\
C$_2$H$_5$CN  & 9  & 147 -- 311 & \uncsci{2.98}{-0.17}{+0.18}{15} & \unc{161.35}{-14.65}{+25.10} & \unc{4.87}{-0.20}{+0.38} & \unc{-3.73}{-0.12}{+0.13}\\
      \hline
      \hline\\[-9pt]
\multicolumn{7}{c}{COMs with $T_\mathrm{ex}$~=~210.7\,K, FWHM~=~5.43\,km\,s$^{-1}$ and $v_\mathrm{LSR}$~=~$-$4.66\,km\,s$^{-1}$}\\
      \hline\\[-9pt]
CH$_3$OH      & 13 & 56 -- 835  & \uncsci{2.04}{-0.03}{+0.03}{18} &  &  & \\
CH$_2$DOH     & 3  & 113 -- 167 & \uncsci{3.40}{-0.03}{+0.03}{17} &  &  & \\
CH$_3$CHO     & 17 & 65 -- 347  & \uncsci{2.17}{-0.05}{+0.05}{16} &  &  & \\
CH$_3$OCH$_3$ & 8  & 81 -- 253  & \uncsci{1.10}{-0.02}{+0.02}{17} &  &  & \\
CH$_3$OCHO    & 7  & 106 -- 312 & \uncsci{7.68}{-0.16}{+0.20}{16} &  &  & \\
C$_2$H$_5$CN  & 9  & 147 -- 311 & \uncsci{2.85}{-0.07}{+0.07}{15} &  &  & \\
      \hline
    \end{tabular}
    \tablefoot{In the case of a single value of $E_\mathrm{up}$ when n$_\mathrm{lines}>1$, it means that those lines are all within 1\,K of each other.\\ \tablefoottext{a}{For this species, $v_\mathrm{LSR}$ was fixed to the value obtained for CH\(_3\)OH.}}
\end{table*}

\subsection{OO~Ser}
Similarly to V346~Nor and Haro~5a~IRS, the source size was not constrained and we assumed the one obtained from the Moment~0 maps, 0.155$''$.
Most of the lines in OO~Ser appear to be made of two components, with the secondary component being blueshifted and $\lesssim$60\% weaker (depending on the line) than the main component.
We attempted to add a second component to the fitting procedure but, because some of the intensities of this possible secondary component are close to the noise level, we could not find consistent results during the runs.
To try to circumvent this issue, we focused on the strongest component.
We also fixed the values of FWHM and $v_\mathrm{LSR}$ to 2.0\,km\,s$^{-1}$ and 9.25\,km\,s$^{-1}$, respectively.
Therefore, $N$ and $T_\mathrm{ex}$ were the only free parameters, covering the same parameter space as the runs for the previous targets.
In a second step, we fitted the remaining COMs with $N$ being the sole free parameter and $T_\mathrm{ex}$ fixed to the median of the values from in the previous run.
The best-fit values and their statistical uncertainties are in \autoref{tab:ooser_bestfit}, and the best-fit spectra are shown in \href{https://zenodo.org/records/14967612}{Appendix F}.
The dominant peaks in the line profiles of OO~Ser appear to be well described with one component.
With the exception of panel $c$ in \href{https://zenodo.org/records/14967612}{Figure F.4}, the residuals between the data and the best-fit models are all within the 3$\sigma$ level.

\begin{table*}
  \caption{Best-fit parameters obtained for OO~Ser assuming a source size of 0.155$''$.\label{tab:ooser_bestfit}}
  \centering
    \begin{tabular}[c]{ccccd{4.3}@{\hskip 0.4in}cd{2.6}}
      \hline\\[-9pt]
      Species & n$_{\mathrm{lines}}$ & $E_\mathrm{up}$ [K] & $N$ [cm$^{-2}$] & \multicolumn{1}{c}{$T_{\mathrm{ex}}$ [K]} & FWHM [km\,s$^{-1}$] & \multicolumn{1}{c}{$v_\mathrm{LSR}$ [km\,s$^{-1}$]}\\
      \hline
      \hline\\[-9pt]
      \multicolumn{7}{c}{COMs with FWHM~=~2.0\,km\,s$^{-1}$ and $v_\mathrm{LSR}$~=~9.25\,km\,s$^{-1}$}\\
      \hline\\[-9pt]
CH$_3$OH       & 8  & 56 -- 746  & \uncsci{2.52}{-0.11}{+0.11}{17} & \unc{158.70}{-5.44}{+9.40}   &  & \\
CH$_3$CHO      & 14 & 65 -- 351  & \uncsci{6.39}{-0.46}{+0.64}{15} & \unc{142.40}{-11.80}{+15.00} &  & \\
CH$_3$OCH$_3$  & 4  & 81         & \uncsci{3.00}{-0.68}{+0.70}{16} & \unc{174.35}{-25.25}{+22.31} &  & \\
CH$_3$OCHO     & 7  & 106 -- 312 & \uncsci{3.89}{-0.18}{+0.29}{16} & \unc{150.90}{-6.30}{+11.80}  &  & \\
CH$_3$COCH$_3$ & 8  & 115        & \uncsci{1.95}{-0.59}{+0.73}{16} & \unc{151.60}{-19.57}{+20.70} &  & \\
      \hline
      \hline\\[-9pt]
\multicolumn{7}{c}{COMs with $T_\mathrm{ex}$~=~151.6\,K, FWHM~=~2.0\,km\,s$^{-1}$ and $v_\mathrm{LSR}$~=~9.25\,km\,s$^{-1}$}\\
      \hline\\[-9pt]
CH$_3$OH       & 8  & 56 -- 746  & \uncsci{2.54}{-0.08}{+0.08}{17} &  &  & \\
CH$_3$CHO      & 14 & 65 -- 351  & \uncsci{6.58}{-0.32}{+0.28}{15} &  &  & \\
CH$_3$OCH$_3$  & 4  & 81         & \uncsci{2.35}{-0.10}{+0.11}{16} &  &  & \\
CH$_3$OCHO     & 7  & 106 -- 312 & \uncsci{3.17}{-0.15}{+0.15}{16} &  &  & \\
CH$_3$COCH$_3$ & 8  & 115        & \uncsci{1.97}{-0.12}{+0.12}{16} &  &  & \\
      \hline
    \end{tabular}
  \tablefoot{In the case of a single value of $E_\mathrm{up}$ when n$_\mathrm{lines}>1$, it means that those lines are all within 1\,K of each other.}
\end{table*}

\subsection{Small molecules}
For the small molecules, we only have one transition of each.
To derive column densities, we consequently need to assume a source size and $T_\mathrm{ex}$.
Thus, we estimated their column densities following two approaches.

First, for each target, we fixed the source size and $T_\mathrm{ex}$ to the values obtained for the COMs, and we let $N$, $v_\mathrm{LSR}$ and FWHM be free parameters.
We then ran an MCMC as for the previous fits.
The best-fit values and uncertainties of all targets are reported in the second, third and fourth columns of \autoref{tab:small}.

Second, we assumed a source size of 2$''$, so that there is no beam dilution, and considered two $T_\mathrm{ex}$ values (50\,K and 300\,K) to provide a range of possible column densities for these molecules.
Here we fixed the $v_\mathrm{LSR}$ and FWHM to the values obtained in the first approach, and computed another set of MCMC runs.
The best-fit values and uncertainties obtained for these two temperatures are shown in columns 5 and 6 of \autoref{tab:small}.

\begin{table*}
  \caption{Best-fit parameters obtained for the small molecules.\label{tab:small}}
  \centering
  \begin{tabular}[c]{cccccccc}
      \hline\\[-9pt]
       & & \multicolumn{3}{c}{$s$ and $T_\mathrm{ex}$ from COMs} & & \multicolumn{2}{c}{$s$=2$''$ and $T_\mathrm{ex}$=50\,K and 300\,K}\\
       \cline{3-5} \cline{7-8}\\[-9pt]
    Species & & $N$ [cm$^{-2}$] & FWHM [km\,s$^{-1}$]& $v_\mathrm{LSR}$ [km\,s$^{-1}$] & & $N_{50\mathrm{K}}$ [cm$^{-2}$] & $N_{300\mathrm{K}}$ [cm$^{-2}$]\\
      \hline
      \hline\\[-9pt]
\multicolumn{8}{c}{L1551~IRS~5~North}\\
      \hline\\[-9pt]
H$_2$S     &  & \uncsci{3.01}{-0.18}{+0.20}{16} & \unc{2.61}{-0.09}{+0.09} & \unc{9.38}{-0.06}{+0.05} &  & \uncsci{1.96}{-0.22}{+0.31}{16} & \uncsci{3.81}{-0.24}{+0.22}{16}\\
DCN        &  & \uncsci{5.00}{-0.35}{+0.34}{14} & \unc{2.61}{-0.12}{+0.15} & \unc{9.68}{-0.06}{+0.06} &  & \uncsci{1.96}{-0.15}{+0.18}{14} & \uncsci{6.53}{-0.35}{+0.42}{14}\\
H$_2$CO    &  & \uncsci{8.83}{-0.62}{+0.62}{16} & \unc{2.30}{-0.10}{+0.11} & \unc{9.43}{-0.06}{+0.06} &  & \uncsci{1.38}{-0.11}{+0.12}{17} & \uncsci{8.54}{-0.53}{+0.61}{16}\\
D$_2$CO    &  & \uncsci{2.77}{-0.20}{+0.18}{15} & \unc{2.90}{-0.19}{+0.14} & \unc{9.35}{-0.09}{+0.07} &  & \uncsci{6.57}{-0.52}{+0.53}{14} & \uncsci{4.91}{-0.33}{+0.32}{15}\\
OC$^{33}$S &  & \uncsci{2.65}{-0.35}{+0.40}{15} & \unc{2.04}{-0.20}{+0.23} & \unc{9.35}{-0.13}{+0.12} &  & \uncsci{2.31}{-0.33}{+0.35}{15} & \uncsci{2.52}{-0.33}{+0.40}{15}\\
SO$_2$     &  & \uncsci{1.24}{-0.14}{+0.16}{16} & \unc{2.42}{-0.26}{+0.30} & \unc{9.10}{-0.12}{+0.13} &  & \uncsci{4.47}{-0.51}{+0.56}{16} & \uncsci{1.04}{-0.11}{+0.12}{16}\\
      \hline
      \hline\\[-9pt]
\multicolumn{8}{c}{L1551~IRS~5~South}\\
      \hline\\[-9pt]
H$_2$S    & & \uncsci{10.00}{-0.78}{+0.76}{16} & \unc{3.08}{-0.14}{+0.17} & \unc{4.12}{-0.06}{+0.07} & & \uncsci{1.10}{-0.09}{+0.09}{16} & \uncsci{2.85}{-0.17}{+0.17}{16} \\
      \hline
      \hline\\[-9pt]
\multicolumn{8}{c}{Haro~5a~IRS}\\
      \hline\\[-9pt]
H$_2$S &  & \uncsci{1.21}{-0.14}{+0.20}{16} & \unc{4.09}{-0.58}{+0.73} & \unc{13.63}{-0.24}{+0.21} &  & \uncsci{1.51}{-0.18}{+0.20}{15} & \uncsci{5.26}{-0.81}{+1.35}{15}\\
DCN    &  & \uncsci{1.59}{-0.28}{+0.32}{14} & \unc{2.96}{-0.41}{+1.23} & \unc{14.00}{-0.36}{+0.27} &  & \uncsci{1.38}{-0.27}{+0.35}{13} & \uncsci{6.29}{-1.00}{+1.19}{13}\\
      \hline
      \hline\\[-9pt]
\multicolumn{8}{c}{V346~Nor}\\
      \hline\\[-9pt]
H$_2$S  &  & \uncsci{7.89}{-0.36}{+0.38}{16} & \unc{5.24}{-0.14}{+0.15} & \unc{-4.33}{-0.09}{+0.09} &  & \uncsci{3.35}{-0.28}{+0.35}{16} & \uncsci{6.56}{-0.30}{+0.29}{16}\\
DCN     &  & \uncsci{2.94}{-0.17}{+0.17}{15} & \unc{6.47}{-0.41}{+0.59} & \unc{-4.14}{-0.19}{+0.19} &  & \uncsci{1.41}{-0.25}{+0.48}{15} & \uncsci{2.57}{-0.17}{+0.17}{15}\\
H$_2$CO &  & \uncsci{1.98}{-0.12}{+0.13}{17} & \unc{6.57}{-0.31}{+0.41} & \unc{-4.52}{-0.16}{+0.19} &  & \uncsci{2.23}{-0.16}{+0.17}{17} & \uncsci{1.53}{-0.10}{+0.10}{17}\\
D$_2$CO &  & \uncsci{1.12}{-0.09}{+0.09}{16} & \unc{5.48}{-0.58}{+0.68} & \unc{-4.54}{-0.21}{+0.18} &  & \uncsci{1.57}{-0.11}{+0.12}{15} & \uncsci{1.25}{-0.09}{+0.09}{16}\\
SO$_2$  &  & \uncsci{9.66}{-0.56}{+0.54}{16} & \unc{6.95}{-0.54}{+0.36} & \unc{-3.72}{-0.15}{+0.26} &  & \uncsci{3.53}{-0.23}{+0.24}{17} & \uncsci{6.70}{-0.36}{+0.35}{16}\\
      \hline
      \hline\\[-9pt]
      \multicolumn{8}{c}{OO~Ser}\\
      \hline\\[-9pt]
H$_2$S  &  & \uncsci{1.14}{-0.11}{+0.12}{16} & \unc{2.79}{-0.23}{+0.32} & \unc{8.32}{-0.12}{+0.08} &  & \uncsci{4.03}{-0.43}{+0.49}{15} & \uncsci{1.25}{-0.13}{+0.14}{16}\\
DCN     &  & \uncsci{3.51}{-0.26}{+0.27}{14} & \unc{2.52}{-0.14}{+0.17} & \unc{9.51}{-0.07}{+0.06} &  & \uncsci{9.41}{-0.82}{+0.81}{13} & \uncsci{3.82}{-0.31}{+0.32}{14}\\
H$_2$CO &  & \uncsci{1.34}{-0.09}{+0.09}{16} & \unc{2.76}{-0.21}{+0.15} & \unc{8.88}{-0.09}{+0.09} &  & \uncsci{1.40}{-0.09}{+0.10}{16} & \uncsci{1.11}{-0.08}{+0.08}{16}\\
D$_2$CO &  & \uncsci{1.30}{-0.17}{+0.16}{15} & \unc{2.72}{-0.37}{+0.73} & \unc{9.72}{-0.11}{+0.10} &  & \uncsci{2.16}{-0.27}{+0.29}{14} & \uncsci{1.82}{-0.22}{+0.22}{15}\\
SO$_2$  &  & \uncsci{1.59}{-0.17}{+0.16}{16} & \unc{2.57}{-0.31}{+0.39} & \unc{9.43}{-0.11}{+0.15} &  & \uncsci{4.62}{-0.54}{+0.59}{16} & \uncsci{1.06}{-0.12}{+0.13}{16}\\
      \hline
    \end{tabular}
    \tablefoot{
 The second, third and fourth columns were obtained using the same source size and $T_\mathrm{ex}$ as for the COMs. The fifth and sixth columns show the best-fit column densities assuming a source size of 2$''$, the FWHM and $v_\mathrm{LSR}$ of the third and fourth columns, and $T_\mathrm{ex}$ values of 50\,K (fifth column) and 300\,K (sixth column).
    }
\end{table*}

\begin{table*}
  \caption{Abundances w.r.t.\ CH\(_3\)OH, \(N\)/\(N_{\mathrm{CH}_3\mathrm{OH}}\). The lines of CH\(_3\)OH are probably optically thick, thus the listed abundances with respect to it are considered as upper limits.\label{tab:abundances}}
  \centering
    \begin{tabular}[c]{cccccccc}
      \hline\\[-8pt]
      Species & \multicolumn{3}{c}{L1551~IRS~5~N} & L1551~IRS~5~S & Haro~5a~IRS & V346~Nor & OO~Ser\\
      \hline\\[-8pt]
      Isotopologue & CH\(_3\)OH\tablefootmark{a} & \(^{13}\)CH\(_3\)OH\tablefootmark{b} & CH\(_3\)\(^{18}\)OH\tablefootmark{c} & CH\(_3\)OH & CH\(_3\)OH & CH\(_3\)OH & CH\(_3\)OH\\
      \hline\\[-8pt]
CH$_3$CHO                   & $<$\sci{5.02}{-3} & \sci{2.38}{-3} & \sci{1.44}{-3} & --                & $<$\sci{1.35}{-1} & $<$\sci{1.06}{-2} & $<$\sci{2.59}{-2}\\
CH$_3$OCH$_3$               & $<$\sci{2.13}{-2} & \sci{1.01}{-2} & \sci{6.13}{-3} & $<$\sci{1.71}{-1} & $<$\sci{5.60}{-1} & $<$\sci{5.39}{-2} & $<$\sci{9.25}{-2}\\
CH$_3$OCHO                  & $<$\sci{3.16}{-2} & \sci{1.49}{-2} & \sci{9.07}{-3} & $<$\sci{1.40}{-1} & $<$\sci{3.75}{-1} & $<$\sci{3.76}{-2} & $<$\sci{1.25}{-1}\\
CH$_3$COCH$_3$              & $<$\sci{6.88}{-3} & \sci{3.25}{-3} & \sci{1.98}{-3} & --                & --                & --                & $<$\sci{7.76}{-2}\\
C$_2$H$_5$OH                & $<$\sci{2.12}{-2} & \sci{1.00}{-2} & \sci{6.09}{-3} & --                & --                & --                & --\\
C$_2$H$_5$CN                & $<$\sci{5.53}{-4} & \sci{2.62}{-4} & \sci{1.59}{-4} & --                & --                & $<$\sci{1.40}{-3} & --\\
CH\(_3\)CN\tablefootmark{d} & $<$\sci{1.16}{-2} & \sci{5.49}{-3} & \sci{3.34}{-3} & $<$\sci{1.00}{-1} & --                &                   & --\\
      \hline
    \end{tabular}
    \tablefoot{
\tablefoottext{a}{Using the column density obtained from the high-excitation fit.}
\tablefoottext{b}{Assuming an isotopic ratio of [$^{12}$C]/[$^{13}$C]~=~68.}
\tablefoottext{c}{Assuming an isotopic ratio of [$^{16}$O]/[$^{18}$O]~=~560.}
\tablefoottext{d}{The column density of CH\(_3\)CN was calculated from our best-fit $N$ of \(^{13}\)CH\(_3\)CN and assuming an isotopic ratio of [$^{12}$C]/[$^{13}$C]~=~68.}
}
\end{table*}

\section{Discussion}\label{sec:discussion}
In this Section we discuss the results for each individual FUor-like target, and explore whether they have the qualities necessary to be defined as a hot corino: compact and warm regions that exhibit COMs.
We complement our discussion by comparing our FUor-like targets with other YSOs.

\subsection{L1551~IRS~5}\label{ss:discussion_l1551irs5}
We found that both disks of L1551~IRS~5 have emission lines from multiple COMs in the inner areas of their disks, and their excitation temperatures are above 100\,K. 
Thus, both disks have the requirements to classify as hot corinos.
\citet{Bianchi2020_MNRAS498L87B} had already classified the Northern disk of L1551~IRS~5 as a hot corino and suggested that the Southern component might also be one.
Here we confirm that both disks have hot corino-like properties.

\citet{Bianchi2020_MNRAS498L87B} ran a non-LTE fitting procedure for L1551~IRS~5 and found a lower limit of \sci{1}{19}\,cm$^{-2}$ for the $N$ of CH\(_3\)OH with a $T_\mathrm{ex}$ of 100\,K, a FWHM of \kms{3.5}, and a source size of 0.15\arcsec.
While our column densities for the Northern disk are in the same order of magnitude, a more quantitative comparison between our results and theirs is not trivial due to the different excitation temperatures, source sizes and FWHMs assumed.
In the case of CH\(_2\)DOH, \citet{Bianchi2020_MNRAS498L87B} fixed the source size to be the same size as for the main isotopologue of methanol, and their best-fit values of $T_\mathrm{ex}$ and $N$ are 88\,K and $\geq$\sci{5}{17}\,cm$^{-2}$, respectively.
Their $T_\mathrm{ex}$ is lower than the value we obtained (210.3\,K) and to the median of all the COMs (177.6\,K). 
Both of our best-fit values of $N$ (\sci{5.25}{18} and \sci{2.81}{18} cm$^{-2}$) agree with their lower limit.
For C\(_2\)H\(_5\)OH,  \citet{Bianchi2020_MNRAS498L87B} fixed the temperature and source size to their values obtained from methanol and obtained a best-fit $N$ that is approximately 2--3 times lower than our values, which is in agreement with our results using a smaller source size (0.08$\arcsec$).
Lastly, their estimate of the column density of CH\(_3\)OCHO is comparable to our best-fit values.

We used the best-fit column densities obtained using the median $T_\mathrm{ex}$ of 177.6\,K to calculate the deuteration ratio, D/H, of methanol for the Northern disk of L1551~IRS~5.
The $N$ ratio between CH\(_2\)DOH (\sci{2.81}{18}\,cm$^{-2}$) and CH\(_3\)OH (\sci{1.60}{19}\,cm$^{-2}$) is \sci{1.8}{-1}, which gives a statistical D/H ratio of CH\(_3\)OH of 5.9\%.
The column densities of CH\(_3\)OH from the $^{13}$C and $^{18}$O isotopologues (\sci{3.38}{19}\,cm$^{-2}$ and \sci{5.57}{19}\,cm$^{-2}$, respectively) lead to isotopic ratios of \sci{8.2}{-2} and \sci{5.1}{-2}, and statistical D/H ratios of 2.8\% and 1.7\%.
These isotopic ratios are one or two orders of magnitude higher than what was found for HDO/H$_2$O \citep[\sci{1.6-2.1}{-3};][]{Andreu2023_AA677L17A}.
However, finding a higher D/H ratio for COMs than for water is common in low-mass protostars \citep[e.g.,][for \object{IRAS~16293-2422}]{Persson2013_AA549L3P,Jorgensen2016_AA595A117J}, which is explained by the earlier formation of water in the star formation process \citep[e.g.,][]{Ceccarelli2014_prplconf859C,Furuya2016_AA586A127F,Jensen2021_AA650A172J}.
Our calculated D/H fractionation is in the same order to what was found for CH\(_2\)DOH in
IRAS~16293~A \citep[2.8\%;][]{Manigand2020_AA635A48M},
IRAS~16293~B \citep[2.4\%;][]{Jorgensen2018_AA620A170J},
\object{IRAS~2A} and \object{IRAS~4A} \citep[5.8\% and 3.7\%, respectively;][]{Taquet2019_AA632A19T},
and \object{Ser-emb~11} \citep[4.0\%;][]{MartinDomenech2021_ApJ923155M}.
At the same time, it is higher than what was found for
SVS13-A \citep[0.05--0.24\%;][]{Bianchi2017_MNRAS4673011B},
\object{HH~212} \citep[0.8\%;][]{Bianchi2017_A606L7B},
\object{L483} \citep[0.8\%;][]{Jacobsen2019_AA629A29J}
and for V883~Ori \citep[0.2--0.7\%;][]{Lee2019NatAs3314L,Yamato2024_AJ16766Y}.

\subsection{Haro~5a~IRS}
The line emission from Haro~5a~IRS appears to be compact (see \autoref{fig:methanol_moments} and \autoref{fig:other_moments}).
Taking this with the presence of four COMs and $T_\mathrm{ex}$ above 100\,K, it suggests that Haro~5a~IRS is also similar to a hot corino.
Haro~5a~IRS has a companion 0.7$''$ to the southwest, and we did not find line emission from it.

Compared to the CH\(_3\)OH maps of the other FUor-like targets (\autoref{fig:methanol_moments}), Haro~5a~IRS is the only one that does not show a disk-like velocity distribution of the gas.
Indeed, only the transitions with $E_{\mathrm{up}}$ values from 165.3\,K to 190.4\,K show emission in more than a few pixels compared to the other three shown transitions.
We think that most of the line emission is hidden by the optically thick dust disk.

\subsection{V346~Nor}
Similar to the other targets, V346~Nor can be considered as a hot corino-like given its compact emission of multiple COMs with $T_\mathrm{ex}$ higher than 100\,K.
We used the best-fit column densities obtained using the median $T_\mathrm{ex}$ of 210.7\,K to calculate the deuteration ratio, D/H, of methanol.
The $N$ ratio between CH\(_2\)DOH (\sci{3.40}{17}\,cm$^{-2}$) and CH\(_3\)OH (\sci{2.04}{18}\,cm$^{-2}$) is \sci{1.7}{-1}, which gives a statistical D/H ratio of CH\(_3\)OH of 5.6\%.
These numbers are the same as those found for L1551~IRS~5~North when using the main methanol isotopologue.
However, the possible underestimation of the column density of CH$_3$OH and the low quality of the CH$_2$DOH model  make this value highly uncertain.

\subsection{OO~Ser}
The COM emission is compact and comparable with the size of the continuum disk.
While the $T_\mathrm{ex}$ values of this target are high enough to be considered a hot corino, its median $T_\mathrm{ex}$ (151.6\,K) is the lowest among the other targets, which have $T_\mathrm{ex}$ values between 177\,K and 225\,K.
The lower temperature might be due to the outburst in this source finishing $\sim$20 years ago.
Indeed, \citet{Molyarova2018_ApJ86646M} showed that after the outburst finishes, different COMs stay in the gas phase for different lengths of time (see their Figure~5).
An extreme example of this is the case of CH$_3$OCHO, which, in some models, had enhanced gas-phase column densities for more than 100~years after the outburst finished.

\subsection{Impact of the chosen pixels}\label{ss:impact_other_pixels}
As discussed in Sect.~\ref{sec:results}, the spectra analyzed in this study are taken at a pixel position which maximizes the number of detected lines and species.
However, the moment maps in \autoref{fig:other_moments} indicate that different COMs could exhibit different spatial distributions, which suggests that abundance ratios relative to CH\(_3\)OH may vary across the disk.
We consequently propose to check it for L1551~IRS~5~North.

For L1551~IRS~5~South, Haro~5a~IRS, V346~Nor, and OO~Ser we cannot explore this as the emission is not extended enough.
At alternative positions located one beam away from the primary pixels, we do not detect most of the COMs.

In contrast, the spectral lines at other positions of L1551~IRS~5~North were strong enough to conduct further analysis.
We examined three additional positions in this FUor, at the northeast (NE), southeast (SE), and southwest (SW) sides of the disk, with our original pixel located on the northwest (NW).
See \autoref{fig:pixels}.
The spectra at these three positions present weaker lines, leading to non-detections of CH\(_3\)OCH\(_3\), CH\(_3\)COCH\(_3\) and CH\(_3\)C\(^{15}\)N.
For the detected COMs at these positions (CH\(_2\)DOH, CH\(_3\)CHO, CH\(_3\)OCHO, C\(_2\)H\(_5\)OH, C\(_2\)H\(_5\)CN, and CH\(_3\)C\(^{15}\)N), we conducted line fitting by fixing the source size to the same size as for the NW position (0.08$''$, see \autoref{tab:north_bestfit}).

We followed the same procedure as in Sect.~\ref{sec:analysis} to obtain the column densities at these alternative positions. 
We used the lines of CH\(_3\)\(^{18}\)OH to estimate the column density of methanol.
We estimated the abundance ratios for the COMs detected at the three additional positions and compared them with those from the original fit.
The results are shown in \autoref{tab:ratios_positions}.

We found that, aside from CH\(_2\)DOH, relative abundances to methanol were consistent across the four positions, within a factor of 3.
For CH\(_2\)DOH, the original position showed higher abundances compared to the three other positions by factors of 3--5, possibly indicating an actual enhancement in the column density of this molecule.
Finally, the D/H statistical ratio at these three alternative positions (0.3--0.5\%) would be comparable to what was found for SVS13-A, HH212, L483 and V883~Ori in Sect.~\ref{ss:discussion_l1551irs5}.

\begin{table}
  \caption{Abundance ratios w.r.t.\ CH\(_3\)OH between different positions in L1551~IRS~5~North.\label{tab:ratios_positions}}
  \centering
    \begin{tabular}[c]{ccccccc}
      \hline\\[-8pt]
      Species & $\dfrac{\mathrm{NW}}{\mathrm{NE}}$ & $\dfrac{\mathrm{NW}}{\mathrm{SE}}$ & $\dfrac{\mathrm{NW}}{\mathrm{SW}}$ & $\dfrac{\mathrm{NE}}{\mathrm{SE}}$ & $\dfrac{\mathrm{NE}}{\mathrm{SW}}$ & $\dfrac{\mathrm{SE}}{\mathrm{SW}}$\\[5pt]
      \hline\\[-8pt]
      CH\(_2\)DOH         & 4.22 & 5.00 & 3.43 & 1.18 & 0.81 & 0.69\\
      CH\(_3\)CHO         & 0.99 & 1.60 & 0.93 & 1.62 & 0.94 & 0.58\\
      CH\(_3\)OCHO        & 1.04 & 0.67 & 0.79 & 0.64 & 0.76 & 1.18\\
      C\(_2\)H\(_5\)OH    & 0.81 & 0.62 & 0.79 & 0.76 & 0.98 & 1.29\\
      C\(_2\)H\(_5\)CN    & 0.53 & 0.38 & 0.46 & 0.71 & 0.87 & 1.23\\
      CH\(_3\)C\(^{15}\)N & 0.68 & 0.67 & 0.68 & 0.97 & 1.00 & 1.02\\
      \hline
    \end{tabular}
\end{table}

\subsection{Comparison with other YSOs}
\citet{Molyarova2018_ApJ86646M} modeled the changes in the chemical composition of a protoplanetary disk due to the effects of a FUor-like luminosity outburst.
They found that some molecules are more sensitive to the outburst than others, meaning that for some of them their abundances w.r.t.\ H$_2$ increase by at least two orders of magnitude.
Four of the COMs we detected are among the ones analyzed by \citet{Molyarova2018_ApJ86646M}: CH\(_3\)OH, CH\(_3\)CHO, CH\(_3\)OCH\(_3\) and CH\(_3\)OCHO.
They found that the abundances w.r.t.\ H$_2$ of these species increase by at least three orders of magnitude during a FUor-type outburst.
Their estimated abundance increases (their Table~1) suggest that for CH\(_3\)CHO, CH\(_3\)OCH\(_3\) and CH\(_3\)OCHO, their abundances w.r.t.\ CH\(_3\)OH should increase one and two orders of magnitude during an outburst.
Furthermore, \citet{Molyarova2018_ApJ86646M} found that CH\(_3\)OH freezes out as soon as the outburst finishes.
However, in their models with medium sized grains the abundance of CH\(_3\)OH w.r.t.\ H$_2$ experiences a second enhancement after the return to quiescence.
The models with large dust grains did not exhibit this behavior.
Their models for CH\(_3\)CHO and CH\(_3\)OCHO show that they have enhanced abundances even 20--300 years after the outburst (depending on the species and the model; see their Figure 5).
Therefore, for large dust grains, the abundances of these two COMs with respect to CH\(_3\)OH would be higher in the post-outburst stage than during the outburst \citep[see also][]{Taquet2016_ApJ82146T}.
In contrast, for medium dust grains, the abundances would be similar during and after the outburst.

\begin{figure*}
  \centering
  \includegraphics[width=\linewidth]{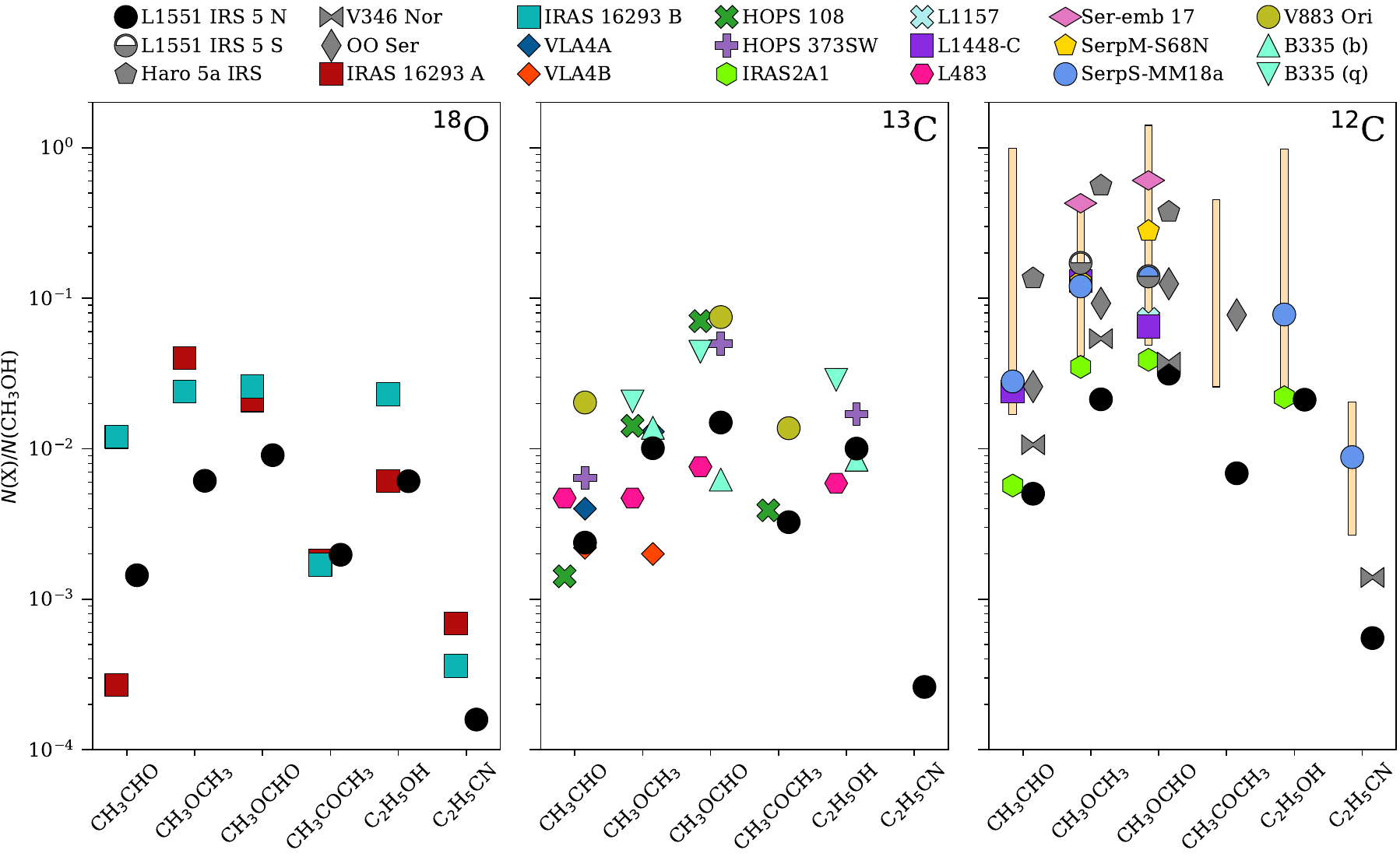}
\caption{Comparison of abundances between our four targets and other young stellar objects. The panels from left to right show the abundance ratios w.r.t.\ methanol based on CH\(_3\)\(^{18}\)OH, \(^{13}\)CH\(_3\)OH, and CH\(_3\)OH respectively. Sources considered to be or have been in outburst (L1551~IRS~5~North, Haro~5a~IRS, V346~Nor, OO~Ser, VLA4A/B, HOPS~373SW, V883~Ori) are slightly shifted to the right, while quiescent sources are shifted to the left. In the case of B335 we plotted two epochs, one from 2014 representing the quiescent state and another from 2016 when the source was bursting. We consider the abundance ratios calculated w.r.t.\ the main isotopologue of CH\(_3\)OH as upper limits.\label{fig:abundances_comparison}}
\end{figure*}

We compare the chemical composition of these FUor-like eruptive young stars with other Class~0/I objects including quiescent sources from the literature by determining the abundance ratios of COMs w.r.t.\ CH$_3$OH.
The goal is to determine whether the presence of bursts could affect the molecular composition and search for differences in the abundances w.r.t.\ methanol that were predicted by \citet{Molyarova2018_ApJ86646M}.
The comparison plot is shown in \autoref{fig:abundances_comparison}.
We limited this comparison only to sources that have been observed with interferometers, as the single-dish observations can cause significant beam-dilution of the emission line fluxes and are more susceptible to envelope or large scale shock contamination.
Our comparison was done against
\object{IRAS~16293--2422} \citep{Jorgensen2016_AA595A117J, Manigand2020_AA635A48M},
\object{V883~Ori} \citep{Lee2019NatAs3314L},
\object{L483} \citep{Jacobsen2019_AA629A29J},
the CALYPSO survey \citep{Belloche2020_AA635A198B},
\object{SVS13-A} \citep{Bianchi2022_ApJ928L3B},
\object{HOPS~108} \citep{Chahine2022_AA657A78C},
the PEACHES survey \citep{Yang2022_ApJ92593Y},
and
\object{HOPS~373SW} \citep{Lee2023_ApJ95643L}.
IRAS~16293--2422, also a protobinary, L1483, HOPS~108, and B335 are Class~0 protostars.
As mentioned in Sect.~\ref{sec:intro}, V883~Ori, SVS13-A, HOPS~373SW, and B335 are outbursting protostars.
The near-infrared spectrum of V883~Ori is similar to the prototypical FUors but its photometric brightening was not detected, thus it is classified as FUor-like \citep{ConnelleyReipurth2018_ApJ861145C}.
SVS13 is a protobinary that went into outburst in 1990 and has been fading since then.
HOPS~373SW began experiencing a powerful outburst in 2020 \citep{Yoon2022_ApJ92960Y}.
B335 began its burst between 2010 and 2012, and it lasted until 2023.
\citet{Lee2025_ApJ978L3L} found that the line intensities of the COMs were higher during the two epochs during the burst and the one at the end of the event compared to the epoch before the burst.
\citet{Lee2025_ApJ978L3L} interpreted this as an indication of an enhancement of the COM column densities due to the sublimation brought by the outburst, and that the COMs had not returned to their pre-burst abundances months after the object returned to quiescence.
The PEACHES survey is composed of 50 Class~0/I sources in Perseus.
From the CALYPSO survey, we included into our analysis only the five sources with estimated abundances (\object{NGC1333-IRAS2A1}, \object{Serps-MM18a}, \object{L1448-C}, \object{SerpM-S68N}, and \object{L1157}), which all happen to be Class~0 objects.

For the four eruptive young stars of this work, we used the abundances calculated w.r.t.\ CH\(_3\)OH.
We also compared the abundances of L1551~IRS~5~North calculated using \(^{13}\)CH\(_3\)OH and CH\(_3\)\(^{18}\)OH. 
In the case of HOPS~108 and the two protostars in IRAS~16293, the abundances were computed using CH\(_3\)\(^{18}\)OH and an isotopic ratio [$^{16}$O]/[$^{18}$O] of 560.
For V883~Ori \citep{Lee2019NatAs3314L}, SVS13-A \citep{Bianchi2022_ApJ928L3B}, and HOPS~373SW \citep{Lee2023_ApJ95643L}, their respective authors calculated their abundances using $^{13}$CH\(_3\)OH and an isotopic ratio [$^{12}$C]/[$^{13}$C] of 60.
We computed the column density ratios of B335 using the \(^{13}\)CH\(_3\)OH column densities reported by \citet{Lee2025_ApJ978L3L}, and an isotopic ratio [$^{12}$C]/[$^{13}$C] of 68.
For the PEACHES survey \citep{Yang2022_ApJ92593Y}, the abundances were calculated w.r.t.\ CH\(_3\)OH.

In \autoref{fig:abundances_comparison}, we see that the abundance ratios of COMs w.r.t.\ CH$_3$OH vary depending on the used  isotopologue. On average, the ratios increase with the abundance of the methanol isotopologue. While it is clearly shown in the COMs abundances of L1551~IRS~5~North for which we also have the $^{13}$C and $^{18}$O isotopologues of methanol, this effect can also be seen along the sources from the literature. This trend strongly suggests that ratios using the main isotopologue of methanol should be considered with extreme caution. Indeed, most of the lines of the CH\(_3\)OH main isotopologue are optically thick. Studies based on the main isotopologue only consequently leads to overestimates of the column density ratios w.r.t.\ CH$_3$OH. 
Comparison using lower abundant isotopologues of CH$_3$OH should be favored when possible. 

In the Northern disk of L1551~IRS~5, we also find that the abundance ratios calculated using \(^{13}\)CH\(_3\)OH are higher than those computed using CH\(_3\)\(^{18}\)OH.
This suggests that opacity effects could also affect the column densities of $^{13}$CH$_3$OH if the assumed isotopic ratios of C and O are correct.
\citet{Bianchi2020_MNRAS498L87B} suggested that the \(^{13}\)CH\(_3\)OH lines could be optically thick in this source too.
So it is possible that it is also the case in other protostars and some abundances of COMs determined relative to \(^{13}\)CH\(_3\)OH in the literature may also represent upper limits.

The range of abundances found using \(^{13}\)CH\(_3\)OH and CH\(_3\)\(^{18}\)OH spans one or two orders of magnitude depending on the species.
The number of sources available for comparison is however very small for a valuable comparison.
Regarding the values of IRAS16293~A taken from \citet{Manigand2020_AA635A48M}, it should be noted that the position chosen to model the species could affect the abundances ratios, as some species (e.g., CH$_3$CHO) seem to be more compact than others.
The abundance ratios given for this source could consequently be different towards a more central position, but blending issues prevent to constrain them precisely.
The low CH$_3$CHO/CH$_3$OH ratio seen in \autoref{fig:abundances_comparison} for IRAS16293~A may then be underestimated.  

\begin{figure*}
  \centering
  \includegraphics[width=\linewidth]{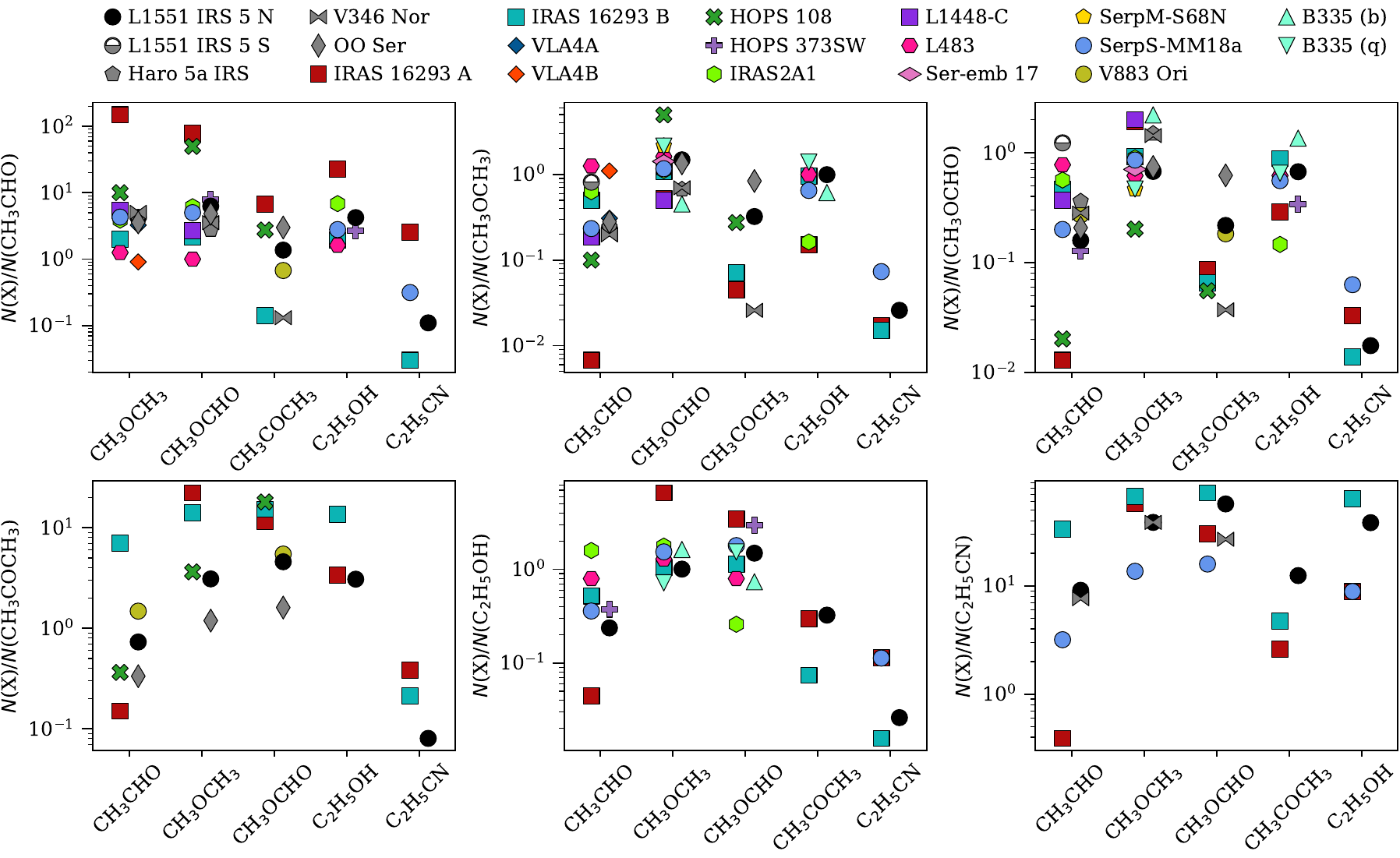}
  \caption{Abundances of the four FUors and other young stellar objects with respect to different COMs. Here the symbols are also slightly shifted to the left or right if the source is considered to be quiescent or outbursting, respectively.\label{fig:com_ratios}}
\end{figure*}
To circumvent the optical thickness issues in CH\(_3\)OH, we also computed abundances with respect to the other COMs.
The results are shown in \autoref{fig:com_ratios}, which do not include L1157 as only CH\(_3\)OH and CH\(_3\)OCHO were detected in it.
These abundances also range between one or two orders of magnitude.
IRAS16293~A is a common outlier when using CH\(_3\)CHO, due to the aforementioned dependence on the position where the spectrum was extracted.
HOPS~108 is also separated from the rest of the sample in a few ratios (CH\(_3\)OCHO/CH\(_3\)CHO, CH\(_3\)OCHO/CH\(_3\)OCH\(_3\)).

Regardless of the COM used to calculate the abundance ratios, no trend is seen between the outbursting and quiescent sources but the sample is too small for any firm conclusion.
In order to properly characterize the diversity of chemical composition in the Class 0/I stages of the star formation process, it is necessary to increase the number of protostars (both quiescent and outbursting) with high-quality measurements of multiple COMs, including rarer isotopologues of CH\(_3\)OH.

\section{Conclusions}\label{sec:conclusions}
We analyzed ALMA Band 6 observations of four FUor-type eruptive young stars, and found emission from multiple COMs and other simple species in each target.
The emission of the COMs is generally compact and comparable to the size of the dust disk.
We used CASSIS to run a line-fitting procedure assuming LTE, obtaining a best-fit column density $N$ and excitation temperature $T_\mathrm{ex}$ for each molecule in each protostar.
Our best-fit resulted in most COMs having $T_\mathrm{ex}$ values above 100\,K, indicating warm emission. Therefore, we propose that these four FUor-type protostars are hot corinos. 

One of these sources, L1551~IRS~5, is a protobinary with COMs emission in both the Northern and Southern disks.
Earlier works have proposed the Northern disk to be the one currently experiencing the outburst, which aligns with our results of it having stronger emission from more COMs than the Southern one, and with higher $T_\mathrm{ex}$ values. 
Haro~5a~IRS has the weakest line emission despite having comparable luminosities and at a comparable distance to OO~Ser.
We expect this to be due to high dust optical depths.
V346~Nor is the source where we detected the second  largest amount of species.
OO~Ser has the lowest $T_\mathrm{ex}$ values among the outbursting sources, but still high enough to be considered as a hot corino. This could be due to the fact that this protostar experienced an 11-year outburst that finished in 2006 \citep{Kospal2007_AA470211K}.

We calculated the abundances of COMs w.r.t.\ CH\(_3\)OH for each target (see \autoref{tab:abundances}). 
In the case of the Northern disk of L1551~IRS~5, several high excitation lines of CH\(_3\)OH are detected.
Their opacities are lower than the ones of the low excitation lines, which are clearly optically thick.
The best-fit $N$ calculated using these high excitation lines is in agreement within a factor 2--3 with the value obtained for \(^{13}\)CH\(_3\)OH and CH\(_3\)\(^{18}\)OH assuming standard isotopic ratios. 
The $^{13}$CH\(_3\)OH lines of L1551~IRS~5~North could be slightly optically thick, if the assumed isotopic ratios are correct.
For the remaining FUors, the CH\(_3\)OH emission lines should be considered as optically thick and the abundance ratios of COMs as upper limits.
Comparisons of COMs abundance ratios w.r.t.\ CH$_3$OH using the rarer $^{18}$O or $^{13}$C isotopologues should be favored.  

The abundance ratios of COMs in L1551~IRS~5 are compared to other values in the literature.
The sample of sources with ratios using the $^{18}$O or $^{13}$C isotopologues of methanol is small, which limits the comparison.
The range of values spans 1--2 orders of magnitude for each species (CH$_3$CHO, CH$_3$OCH$_3$, CH$_3$OCHO, CH$_3$COCH$_3$, C$_2$H$_5$OH, C$_2$H$_5$CN).
No trend is seen in the outbursting sources vs the more quiescent hot corinos.

COM detection does not only occur in outbursting stars.
However, it is clear that FUor-type eruptive young stars could be used as experiments to study the chemistry of protostars that otherwise would not be luminous enough to have significant COM detection.
Comparing different sources observed using different instrumentation or even spectral configurations is not trivial due to a difference in optical depths or range of $E_\mathrm{up}$ values analyzed.
Therefore, surveys with deep integrations and broad spectral bandwidth that covers multiple transitions from different isotopologues is the best option to study the chemistry of protostars and to carry out comparisons between them.

\section*{Data availability}
Appendices are available in \url{https://zenodo.org/records/14967612}.

\begin{acknowledgements}
We are grateful to the anonymous referee for their constructive comments, which significantly enhanced the quality of this paper.
This study is part of a project that has received funding from the European Research Council (ERC) under the European Union’s Horizon 2020 research and innovation programme (Grant agreements no.\ 949278, Chemtrip, and no.\ 716155, SACCRED).
This paper makes use of the following ALMA data: ADS/JAO.ALMA\#2016.1.00209.S.
ALMA is a partnership of ESO (representing its member states), NSF (USA) and NINS (Japan), together with NRC (Canada) and NSC and ASIAA (Taiwan) and KASI (Republic of Korea), in cooperation with the Republic of Chile.
The Joint ALMA Observatory is operated by ESO, AUI/NRAO and NAOJ\@.
On behalf of the SACCRED project we thank for the usage of MTA Cloud (\url{https://cloud.mta.hu/}) that significantly helped us achieving the results published in this paper.
\end{acknowledgements}

\bibliographystyle{aa}
\bibliography{fcsm_fuors_hot_corinos}

\end{document}